\documentclass{article}
\usepackage[T1]{fontenc}
\usepackage{wrapfig}
\usepackage[portuges,english]{babel}
\usepackage{amssymb,latexsym,amsmath,color,mathrsfs,ifsym,graphics,stmaryrd} 
\usepackage[colorlinks,linkcolor=blue,urlcolor=blue,citecolor=black,
plainpages=false,pdfpagelabels,breaklinks]{hyperref}
\usepackage{dialogue}
\usepackage{graphicx}
\title{{\sc Understanding Quantum Mechanics\\
(Beyond Metaphysical Dogmatism and Naive Empiricism)}}

\author{{\sc Christian de Ronde}\thanks{Fellow Researcher of the Consejo
Nacional de Investigaciones Cient\'{\i}ficas y T\'ecnicas. E-mail: cderonde@vub.ac.be}}
\date{}

\usepackage[margin=2.08cm]{geometry}

\begin{document}
\maketitle

\begin{center}
\begin{small}
Philosophy Institute Dr. A. Korn, Buenos Aires University - CONICET\\ 
Engineering Institute - National University Arturo Jauretche, Argentina.\\
Federal University of Santa Catarina, Brazil.\\ 
Center Leo Apostel for Interdisciplinary Studies, Brussels Free University, Belgium.\\
\end{small}
\end{center}

\bigskip

\begin{abstract}
\noindent Quantum Mechanics (QM) has faced deep controversies and debates since its origin when Werner Heisenberg proposed the first mathematical formalism capable to operationally account for what had been recently discovered as the new field of quantum phenomena. Today, even though we have reached a standardized version of QM  which is taught in Universities all around the world, there is still no consensus regarding the conceptual reference of the theory and, if or if not, it can refer to something beyond measurement outcomes. In this work we will argue that the reason behind the impossibility to reach a meaningful answer to this question is strictly related to the 20th Century Bohrian-positivist re-foundation of physics which is responsible for having introduced within the theory of quanta a harmful combination of metaphysical dogmatism and naive empiricism. We will also argue that the possibility of understanding QM is at plain sight, given we return to the original framework of physics in which the meaning of understanding has always been clear.
\medskip\\
\noindent \textbf{Key-words}: Interpretation, explanation, representation, quantum theory.
\end{abstract}

\renewenvironment{enumerate}{\begin{list}{}{\rm \labelwidth 0mm
\leftmargin 0mm}} {\end{list}}

\newcommand{\ita}{\textit}
\newcommand{\mcal}{\mathcal}
\newcommand{\mfrak}{\mathfrak}
\newcommand{\mbb}{\mathbb}
\newcommand{\mrm}{\mathrm}
\newcommand{\msf}{\mathsf}
\newcommand{\mscr}{\mathscr}
\newcommand{\lra}{\leftrightarrow}
\renewenvironment{enumerate}{\begin{list}{}{\rm \labelwidth 0mm
\leftmargin 5mm}} {\end{list}}

\newtheorem{dfn}{\sc{Definition}}[section]
\newtheorem{thm}{\sc{Theorem}}[section]
\newtheorem{lem}{\sc{Lemma}}[section]
\newtheorem{cor}[thm]{\sc{Corollary}}
\newcommand{\Proof}{\textit{Proof:} \,}
\newcommand{\cqd}{{\rule{.70ex}{2ex}} \medskip}

\bigskip

\bigskip

\bigskip

\bigskip

\bigskip

\begin{flushright}
{\small {\bf Dogma:} {\it A fixed, especially religious,\\
belief or set of beliefs that people are\\ 
expected to accept without any doubts.}\footnote{Definition taken from the Cambridge Dictionary.}} 
\end{flushright}

\section{``Nobody Understands Quantum Mechanics''}

During his lecture at Cornell University in 1964, the U.S. physicist Richard Feynman famously declared to a packed audience of young undergraduate students: ``I think I can safely say that nobody understands quantum mechanics.'' In a kind of sarcastic tone Feynman continued his speech by warning students about the dangers of actually trying to understand the theory of quanta:  
\begin{quotation}
\noindent {\small ``So do not take the lecture too seriously, feeling that you really have to understand in terms of some model what I am going to describe, but just relax and enjoy it. I am going to tell you what nature behaves like. If you will simply admit that maybe she does behave like this, you will find her a delightful, entrancing thing. Do not keep saying to yourself, if you can possibly avoid it, `But how can it be like that?' because you will get `down the drain', into a blind alley from which nobody has yet escaped. Nobody knows how it can be like that.'' 
\cite[p. 129]{Feynman67}} 
\end{quotation}
Feynman, the most famous U.S. physicist that has ever lived, was actually advising young students eager to learn about nature and science not to engage in any attempt to try to understand the most important theory of the 20th Century. Instead, he recommended them to passively accept, without asking further questions, the way that ``nature behaves''. These words, spoken by Feynman, can be only understood in the context of the post-war anti-realist re-foundation of physics. As described by Olival Freire Jr.: 
\begin{quotation}
\noindent {\small ``In the US, which after the Second World War became the central stage of research in physics in the West, the discussions about the interpretation of quantum mechanics had never been very popular. A common academic policy was to gather theoreticians and experimentalists in order to favour experiments and concrete applications, rather than abstract speculations (Schweber 1986). This practical attitude was further increased by the impressive development of physics between the 1930s and the 1950s, driven on the one hand by the need to apply the new quantum theory to a wide range of atomic and subatomic phenomena, and on the other hand by the pursuit of military goals. As pointed out by Kaiser (2002, pp. 154-156), `the pedagogical requirements entailed by the sudden exponential growth in graduate student numbers during the cold war reinforced a particular instrumentalist approach to physics'.'' \cite[pp. 77-78]{Freire15}} 
\end{quotation}
The orthodox understanding of physics became completely detached from metaphysical and philosophical considerations and grounded instead in a down to earth pragmatic approach. At the beginning of the 1960s, Karl Popper \cite{Popper63} had already described the culmination of this radical transition between realism and anti-realism: ``Today the view of physical science founded by Osiander, Cardinal Bellarmino, and Bishop Berkeley, has won the battle without another shot being fired. Without any further debate over the philosophical issue, without producing any new argument, the {\it instrumentalist} view (as I shall call it) has become an accepted dogma. It may well now be called the `official view' of physical theory since it is accepted by most of our leading theorists of physics (although neither by Einstein nor by Schr\"odinger). And it has become part of the current teaching of physics.'' In the U.S., the philosopher John Dewey developed what he called {\it instrumentalism}, a philosophy that could be regarded as the natural extension of both pragmatism and empirical positivism. While pragmatism sustained that the value of an idea is determined by its usefulness; instrumentalism, by explicitly rejecting the need of any metaphysical fundament, was ready to take a step further and claim explicitly that the question regarding the {\it reference} of theories was simply meaningless:  {\bf Scientific theories do not make reference to an underlying reality.} Thus, there would exist no reason to derive a ``model'' that would explain how and why ``nature behaves'' the way it does. Instead, theories should be regarded as ``tools'' that allow us, humans, to compute measurement outcomes. Period. As remarked by Popper, the {\it Manhattan Project} ---in which Feynman also took part--- had also played an essential role in fundamenting the new doctrine: ``Instead of results due to the principle of complementarity other and more practical results of atomic theory were obtained, some of them with a big bang. No doubt physicists were perfectly right in interpreting these successful applications as corroborating their theories. But strangely enough they took them as confirming the instrumentalist creed.'' Knowing or not, Feynman was already part of a movement which after many centuries of countless battles had finally defeated realism and imposed a new (anti-realist) understanding of physics. Willingly or not, Feynman supported instrumentalism in two different ways. First, going explicitly against philosophical and metaphysical considerations. An exposition of his hostility against philosophy can be found in the first volume of his famous {\it Lectures on Physics} \cite[p. 8.2]{Feynman63} where ---making fun of philosophers--- he wrote the following: ``We can't define anything precisely. If we attempt to, we get into that paralysis of thought that comes to philosophers who sit opposite each other, one saying to the other: `you don't know what you are talking about!'. The second one says: `what do you mean by talking? What do you mean by you? What do you mean by know?'.'' Secondly, Feynman expressed the idea that physical theories are just models which are mainly created in order to ``solve problems''. His down-to-earth perspective about physics is clearly exposed in his autobiographical book of anecdotes, {\it Surely you're joking Mr. Feynman!}, where there are many examples of his practical approach to physics. In particular, it is interesting to notice the way in which he described the situation of physicists in the late 1930s, just before the war \cite[p.53]{Feynman85}: ``At the time nobody knew what a physicist even was, and there weren't any positions in industry for physicists. Engineers, OK; but physicists ---{\it nobody knew how to use them.} It's interesting that very soon, after the war, it was the exact opposite: people wanted physicists everywhere.'' This idea, according to which physicists provide solutions to problems through the construction of mathematical models, was an essential part of the instrumentalist creed. 

Since Feynman declared more than half a century ago that ``nobody understands quantum mechanics'' there has been little  progress about this matter. In fact, the idea that it is simply impossible to understand QM has become completely sedimented in physics. Today ---many decades after Feynman's lecture--- all living generations of physicists, all around the world, have been trained in an instrumentalist fashion not to make nasty questions about the meaning of quantum reality. Today, instrumentalism is stronger than ever and QM is taught to students in Universities as a ``recipe'' that predicts how ``nature behaves'' ---even though nobody can explain how or why it does so. As very clearly described by the U.S. philosopher of physics, and former physicist, Tim Maudlin: 
\begin{quotation}
\noindent {\small ``What is presented in the average physics textbook, what students learn and researchers use, turns out not to be a precise physical theory at all. It is rather a very effective and accurate recipe for making certain sorts of predictions. What physics students learn is how to use the recipe. For all practical purposes, when designing microchips and predicting the outcomes of experiments, this ability suffices. But if a physics student happens to be unsatisfied with just learning these mathematical techniques for making predictions and asks instead what the theory claims about the physical world, she or he is likely to be met with a canonical response: Shut up and calculate!'' \cite[pp. 2-3]{Maudlin19}} 
\end{quotation} 


 

\section{Understanding Within Physics: The One and the Many}

History is essential to know and understand ourselves. The bounds with our past are always kernel to understand our present and decide about our future. The same happens with any community or traditions of thought. In this particular respect, it is of outmost importance to recognize that the origin of physics goes back to ancient Greece. As remarked by Jean-Pierre Vernant \cite[p. 400]{Vernant06}: ``Everything began in the early sixth century BCE, in the Greek city of Miletos, on the coast of Asia Minor, where the Ionians had established rich and prosperous colonies. Within fifty years, three men ---Thales, Anaximander, and Anaximenes--- appeared in quick succession.'' The fascinating work of these first philosophers was described by Werner Heisenberg with great mastery in his book {\it Physics and Philosophy} \cite[p. 27]{Heis58}: 
\begin{quotation} 
\noindent {\small ``[T]he elaboration of the concepts of Matter, Being and Becoming [...] started in the sixth century BC with Thales, the founder of the Milesian school, to whom Aristotle ascribes the statement: `Water is the material cause of all things.' This statement, strange as it looks to us, expresses, as Nietzsche has pointed out, three fundamental ideas of philosophy. First, the question as to the material cause of all things; second, the demand that this question be answered in conformity with reason, without resort to myths or mysticism; third, the postulate that ultimately it must be possible to reduce everything to one principle. Thales' statement was the first expression of the idea of a fundamental substance, of which all other things were transient forms. [...] The idea of the fundamental substance was then carried further by Anaximander, who was a pupil of Thales and lived in the same town. Anaximander denied the fundamental substance to be water or any of the known substances. He taught that the primary substance was infinite, eternal and ageless and that it encompassed the world. This primary substance is transformed into the various substances with which we are familiar. Theophrastus quotes from Anaximander: `Into that from which things take their rise they pass away once more, as is ordained, for they make reparation and satisfaction to one another for their injustice according to the ordering of time.' In this philosophy the antithesis of Being and Becoming plays the fundamental role. [...] The third of the Milesian philosophers, Anaximenes, an associate of Anaximander, taught that air was the primary substance. `Just as our soul, being air, holds us together, so do breath and air encompass the whole world.' Anaximenes introduced into the Milesian philosophy the idea that the process of condensation or rarefaction causes the change of the primary substance into the other substances. The condensation of water vapor into clouds was an obvious example, and of course the difference between water vapor and air was not known at that time.'' \cite[p27-29]{Heis58}} 
\end{quotation}
As Vernant continues to make explicitly clear, the revolution that begun with the Milesians in Greece was one of enormous proportions. One that determined the origin of science itself. 
\begin{quotation} 
\noindent {\small ``[Before the Milesians,] [e]ducation was based not on reading written texts but on listening to poetic songs transmitted from generation to generation. [...] These songs contained everything a Greek had to know about man and his past ---the exploits of heroes long past; about the gods, their families, and their genealogies, about the world, its form, and its origin. In this respect, the work of the Milesians is indeed a radical innovation. Neither singers nor poets nor storytellers, they express themselves in prose, in written texts whose aim is not to unravel a narrative thread in the long line of a tradition but to present an explanatory theory concerning certain natural phenomena and the organization of the cosmos. In this shift from the oral to the written, from the poetic song to prose, from narration to explanation, the change of register corresponds to an entirely new type of investigation ---new both in terms of its object (nature, {\it physis}) and in terms of the entirely positive form of thought manifested in it.'' \cite[p. 402]{Vernant06}} 
\end{quotation}

Physics talks about {\it physis} ---also translated as `reality' or `nature'. That is why it was called `physics'. In general terms, one might consider the main idea constituting and characterizing both physics and philosophy as a common attempt to capture the famous interplay between the One and the Many. As Heisenberg \cite[p. 29]{Heis58} would reflect: ``Looking back to the development of Greek philosophy up to this point one realizes that it has been borne from the beginning to this stage by the tension between the One and the Many. For our senses the world consists of an infinite variety of things and events, colors and sounds. But in order to understand it we have to introduce some kind of order, and order means to recognize what is equal, it means some sort of unity.'' Indeed, {\it physis}, what the first philosophers considered the fundament of the whole existence, was seen as possessing an internal order, what the Greeks called ---after Heraclitus--- a {\it logos}. This order could be actually known through the development of {\it theories}, grounding the {\it praxis} of physics in the creation of unified, consistent and coherent schemes of thought that would explain phenomena. It is mainly this possibility brought in by {\it theories} to relate, on the one hand, to {\it physis} (the One), and on the other hand, to the multiplicity of phenomena (the Many), which marks the characteristic feature of scientific understanding. While `The One' implies a unified theoretical standpoint of analysis grounded in {\it physis}, `The Many' relates to the multiplicity of different phenomena and observations. In this way, theories became the bridge between the whole of nature and experience. As a young Pauli would explain to Heisenberg during a conversation in 1921: 
\begin{quotation} 
\noindent {\small  ``knowledge cannot be gained by understanding an isolated phenomenon or a single group of phenomena, even if one discovers some order in them. It comes from the recognition that a wealth of experiential facts are interconnected and can therefore be reduced to a common principle. [...] `Understanding' probably means nothing more than having whatever ideas and concepts are needed to recognize that a great many different phenomena are part of coherent whole. Our mind becomes less puzzled once we have recognized that a special, apparently confused situation is merely a special case of something wider, that as a result it can be formulated much more simply. The reduction of a colorful variety of phenomena to a general and simple principle, or, as the Greeks would have put it, the reduction of the many to the one, is precisely what we mean by `understanding'. The ability to predict is often the consequence of understanding, of having the right concepts, but is not identical with `understanding'.'' \cite[p. 63]{Heis71}} 
\end{quotation}

This is the way in which physicists and philosophers engaged in the wildest enterprise that humanity had ever before witnessed: the understanding of reality. The search for consistent and coherent explanations united the Greeks in a common adventure. As described by Schr\"odinger in his beautiful book, {\it Nature and the Greeks}: 
\begin{quotation} 
\noindent {\small ``[T]he true subject [of the ancient Greeks] was essentially one, and that important conclusions reached about any part of it could, and as a rule would, bear on almost every other part. The idea of delimitation in water-tight compartments had not yet sprung up. [...] To put it dramatically: one can imagine a scholar of the young School of Athens paying a holiday visit to Abdera (with due caution to keep it secret from his Master), and on being received by the wise, far-travelled and world-famous old gentleman Democritus, asking him questions on the atoms, on the shape of the earth, on moral conduct, God, and the immortality of the soul ---without being repudiated on any of these points. Can you easily imagine such a motley conversation between a student and his teacher in our days?'' \cite[pp. 14-15]{Schr54}} 
\end{quotation}

Since everyone could ---in principle--- learn about a reality that was not ruled by the wishes and desires of the Gods, physics ---which was also concomitant with the creation of Greek democracy--- soon became the most powerful enemy of the authority based knowledge that prevailed at the time. The many `chosen figures' like kings, mediums, priests and preachers, who had justified their ruling in societies in a preferred access to `the divine' had to give up ---at least--- part of their power to scientific understanding. For the first time mystical and religious doctrines had to confront a powerful rival. But this division ---as we already mentioned--- was not only part of theoretical thinking, it was also related to the creation by the Greeks of a new form of society: democracy. As Moses Finley \cite[pp. 13-14]{Schr54} explains: ``It was the Greeks, after all, who discovered not only democracy but also politics, the art of reaching decisions by public discussion and then of obeying those decisions as a necessary condition of civilized society.'' This new political {\it praxis}, as explained by Vernant, was also intrinsically related to the method of science itself: 
\begin{quotation} 
\noindent {\small ``Taken out of the realm of the secret, the physicist's {\it theoria} thus becomes a subject of debate; it is obliged to justify itself; it must account for its claims and lay itself open to critique and controversy. The rules of the political game ---open discussion, contentious debate, the confrontation of opposite arguments--- become the rules of the intellectual game. Alongside religious revelation, which, in the form of mystery, remains the privilege of a restricted circle of initiates, and alongside the mass of common beliefs that everyone shares without question, a new notion of truth takes shape and is affirmed: open truth, accessible to all, and justified by its own demonstrative force.'' \cite[p. 405]{Vernant06}} 
\end{quotation}
Like democracy, both physics and philosophy flourished in the beautiful islands of the Aegean sea turning its capital, Athens, into the center of the Ancient world. This was until, a few centuries later, the first form of anti-realism arrived to Athens. 

During the 5th. Century B.C., a new group that would became to be known as `sophists' would produced the first assault against realism, fighting the idea that theoretical knowledge about {\it physis} (or reality) was actually possible. Going back to its original meaning of {\it logos} ---which philosophers had related to {\it physis}--- as discourse, sophists realized that its knowledge had an important pragmatic application which could be sold in the streets of Athens, namely, to win debates in the Agora. Grounding their criticisms in a skeptic position sophists produced a radical attempt to change the perspective, from {\it physis} to the {\it subject}, and from {\it theory} to {\it empirical perception}. Protagoras argued ---in a phrase that would become famously used by positivists of the {\it Vienna Circle}--- that: ``Man is the measure of all things, of the things that are, that they are, of the things that are not, that they are not'' [DK 80B1]. According to this exposition, there is no such thing as `a reality of things' ---or at least, we are not able to grasp it---, we can only refer to our own perception. Things do not have a reality independent of subjects; and even in the case such a reality would exist we would not be able to know it. We, individuals, have only a {\it relative} access to things which is dependent on our own personal experience. From this sceptic standpoint, sophists criticized the physical idea of theoretical knowledge: realists ---namely, physicists and philosophers--- are too naive, they believe they can access reality, but even if such thing would exist no one would be able to do so.\footnote{In this respect, it is of outmost importance to point out that realists (i.e., physicists and philosophers) never claimed to describe {\it reality as it is}. In fact, already the first philosophers showed a clear recognition of the problem implied by such a niave idea. There is an affinity between the {\it logos} of men and the {\it logos} of {\it physis}. However, it is a difficult task to expose the true {\it logos} since, as remarked by Heraclitus, ``{\it physis} loves to hide.'' [f. 123 DK]. Doing so requires hard work and sensibility, but ---following Heraclitus--- the latter can be revealed in the former. In a particular {\it logos} one can ``listen'' something that exceeds it, that is not only that personal discourse but the logos of {\it physis}: ``Listening not to me but to the {\it logos} it is wise to agree that all things are one'' [f. 50 DK]. We are thus able to represent {\it physis}, to exhibit its {\it logos}, but this representation should be not confused mirroring reality.} 

It was necessary the strength of both Plato and Aristotle to control the sophist attack and return to the realist path through the development of what would be called later on `metaphysics', namely, the systematic construction of a conceptual architectonic derived from first principles. First Plato, and then Aristotle, would produce the first systems of concepts capable to account for the One and the Many, giving us the possibility to understand what is {\it the same} within the repetition of {\it difference}. In {\it Sophist} Plato goes on to re-define his {\it Theory of Ideas} in terms of {\it being} as {\it dunamis} ``And I hold that the definition of being is simply {\it dunamis}'' [{\it Sophist}, 247e]. The fundamental reality of everything is, for Plato, {\it dunamis} ---also translated as {\it potency} (especially when it is Aristotle who uses it), possibility, capacity or power. N\'estor Cordero, in a commentary to the {\it Sophist} \cite[p. 144]{Cordero14}, describes it as ``the capacity of an entity, any entity, for relating with another (either affecting or being affected) and for that reason a few pages later Plato replaces `acting' and `suffering' by a single verb, `communicating', and he talks about {\it dunamis koinonas}, `the possibility or potentiality of communication' [{\it Op. cit.}, 251e]. And since being is communicating, something that doesn't communicate doesn't exist. (...) Plato assimilates this potentiality to the fact of being and he gives precisions about it: it is the possibility of communicating, that is, (...) to produce reciprocal bonds''. This being in everything that is, it is {\it dunamis} of affecting and being affected, of interacting, it is an inherent tendency towards relation, the participation of a universal relationability (see for a detailed discussion \cite{deRondeFM19}). Aristotle will also develop a metaphysical scheme in order to describe reality. This representation will be multiple right from the start. As he will argue: ``Being is said in many ways''. Indeed, it is said as {\it potential}, but it is also said as {\it actual}. While the potential realm of existence will be conceived by Aristotle ---following his own interpretation of the philosophical scheme of Heraclitus---  as an indetermined and contradictory realm of existence, he will characterize the actual realm in terms of the metaphysical principles of {\it existence}, {\it non-contradiction} and {\it identity}. In this way Aristotle will provide the very first definition of the concept of {\it entity}. According to him an entity is always partly potential and partly actual, and it is through its actualization ---the process going from the potential to the actual--- that the problem of motion and knowledge finds a common solution. It will be this same notion of entity, detached from its potential existence, which will survive in modern times as characterizing a physical objects. 

Since the ancient Greeks and for almost two millennia, in the Western world the works of Plato and Aristotle will be examined, discussed and questioned ---many times under the mask of religious debates. In the context of physics there will be also a continuous increase in the conceptual and mathematical levels of abstraction. And yet, the important breakthrough will have to wait until the 16th and 17th centuries. It is with the creation of infinitesimal calculus ---developed by Newton and Leibniz--- that mathematics will be finally able to capture, through a rigorous definition of the {\it continuum}, two of the main concepts discussed by the Greeks, namely, {\it space} and {\it time}. Even more importantly, the Greek relationship between the One and the Many will also find a rigorous mathematical representation. All these ideas will find a conclusive exposition in a new metaphysical architectonic due to physicist and philosopher Immanuel Kant.


\section{The 17th Century Modern Re-Foundation of Physics}

Science is not the accomplishment of single individuals, it is a result of team work. Individuals can develop scientific knowledge only with the aid of their communities. It is only in open and prosperous economic societies that individuals can follow the path of scientific research. It is certainly true that it was during the 17th Century that the first closed physical theory was synthesized by Isaac Newton. However, this marvelous creation cannot be regarded as the accomplishment of a single man. Classical mechanics was in fact the result of the work of many different individuals, through many centuries, most of them supported by their communities, all of them following the path laid down by the Greek scientific tradition. It was of course the Greeks who had created the basic structure of {\it theories}, and it was also them who had linked physical concepts with mathematical knowledge. Physical theories were both formal and conceptual right from the start. The first philosophers considered the elements constituting {\it physis} in relation to geometrical figures and it is widely acknowledged that the mathematical investigations of Pythagoras of Samos played an essential role in Plato's metaphysics. Euclides' famous book, {\it Elements}, produced a geometrical representation of 3-dimensional space that survived untouched until the end of the 19th Century and Aristotle's metaphysics imposed the creation of classical logic, a scheme which survived up to our days. While from a mathematical perspective, Zeno's famous analysis of motion and the continuum can be regarded as the starting point of the development of infinitesimal calculus by Newton and Leibniz, the picture of the Universe developed by classical mechanics was the conclusion of the path laid down by the atomist metaphysics created by Leucippus and Democritus.\footnote{This doctrine, created by Democritus and Leucipus, argued that the elements constituting {\it physis} where {\it being} and {\it not-being}, which in term they interpreted as `the full' and `the void'. Contrary to Parmenides, they conceived the existence of a non-being which is the contrary of being. In their minds, there is void within nature. Being, or the full, consisted to them of indivisible bodies, indivisible fragments of mass with a minimum size. Simple bodies. They used an adjective to describe these bodies: {\it \'atomos}, which means, literally, ``not divided''. That adjective became an -ism and this school was called ``atomism''.} But all these developments did not end with the Greeks. Medieval philosophers like Augustine of Hippo and Thomas of Aquinas made monumental breakthroughs in the philosophy of time. Nicholas of Cusa went on to develop Aristotelian metaphysics and the potential realm of existence. As remarkably exposed in Alistair Crombie's book, {\it Augustine to Galileo} \cite[p. xi]{Crombie53}, there obviously exists an ``essential continuity of the Western scientific tradition from Greek times to the 17th century and, therefore, to our own day.''

The specific revolution that took place with classical mechanics, which was also a revolution regarding the notions of space and time, had an enormous impact in the understanding of science. And it is only by understanding this essential influence that we can begin to grasp the deep re-foundation that took place in both physics and philosophy during the 17th Century. In the 2nd Century A.D., Claudius Ptolemaeus was able to construct a mathematical model which allowed to predict the movement of the planets with fantastic accuracy. Following Aristotle, the Earth was placed in the center of the Universe and the sun together with the rest of the planets were pictured as orbiting it, following the perfection of circles. It was not until the 16th Century that Nicolas Copernicus was able to reframe this model by placing the sun ---instead of the Earth--- in the center of the Universe. But it was only with Newton that the final synthesis was reached through a unified, consistent and coherent theoretical representation of nature. Since Aristotle, space had been understood as a {\it place}, a {\it topos}, and time as characterizing the way in which things {\it changed}. Following this understanding, a quite obvious distinction was drawn between the stars in the sky which, when observed, looked as fixed static entities, and what happened in the Earth, a place were motion and change ruled everything. After two millennia in the development of Western thought, classical mechanics was finally able to unite both realms. But in order to do so, space and time would become the ``containers'' of the physical world itself (see \cite{Kuhn57}) leaving no space for the potential realm. While for Aristotle ``being is said in many ways'', potential and actual; for Newton the physical world would become restricted to an actualist space-time representation. By restricting {\it physis} to this specific space-time representation, there was now place for the consideration of new realms of existence which essentially differed from (physical) reality. 

It was in the first half of the 17th Century that Ren\'e Descartes would complete the division between matter and mind, between soul and body. As remarked by Heisenberg, this essential separation between {\it res cogitans} and  {\it res extensa}, would also have deep consequences for the division between physics and philosophy: 
\begin{quotation} 
\noindent {\small ``The great development of natural science since the sixteenth and seventeenth centuries was preceded and accompanied by a development of philosophical ideas which were closely connected with the fundamental concepts of science. It may therefore be instructive to comment on these ideas from the position that has finally been reached by modern science in our time.

The first great philosopher of this new period of science was Ren\'e Descartes who lived in the first half of the seventeenth century. Those of his ideas that are most important for the development of scientific thinking are contained in his {\it Discourse on Method}. On the basis of doubt and logical reasoning he tries to find a completely new and as he thinks solid ground for a philosophical system. He does not accept revelation as such a basis nor does he want to accept uncritically what is perceived by the senses. So he starts with his method of doubt. He casts his doubt upon that which our senses tell us about the results of our reasoning and finally he arrives at his famous sentence: {\it `cogito ergo sum.'} I cannot doubt my existence since it follows from the fact that I am thinking. After establishing the existence of the I in this way he proceeds to prove the existence of God essentially on the lines of scholastic philosophy. Finally the existence of the world follows from the fact that God had given me a strong inclination to believe in the existence of the world, and it is simply impossible that God should have deceived me.

This basis of the philosophy of Descartes is radically different from that of the ancient Greek philosophers. Here the starting point is not a fundamental principle or substance, but the attempt of a fundamental knowledge. And Descartes realizes that what we know about our mind is more certain than what we know about the outer world. But already his starting point with the `triangle' God-World-I simplifies in a dangerous way the basis for further reasoning. The division between matter and mind or between soul and body, which had started in Plato's philosophy, is now complete. God is separated both from the I and from the world. God in fact is raised so high above the world and men that He finally appears in the philosophy of Descartes only as a common point of reference that establishes the relation between the I and the world.

While ancient Greek philosophy had tried to find order in the infinite variety of things and events by looking for some fundamental unifying principle, Descartes tries to establish the order through some fundamental division. But the three parts which result from the division lose some of their essence when any one part is considered as separated from the other two parts. If one uses the fundamental concepts of Descartes at all, it is essential that God is in the world and in the I and it is also essential that the I cannot be really separated from the world. Of course Descartes knew the undisputable necessity of the connection, but philosophy and natural science in the following period developed on the basis of the polarity between the {\it `res cogitans'} and the {\it `res extensa,'} and natural science concentrated its interest on the {\it `res extensa.'} The influence of the Cartesian division on human thought in the following centuries can hardly be overestimated, but it is just this division which we have to criticize later from the development of physics in our time.'' \cite[pp. 41-42]{Heis58}} 
\end{quotation}
This separation allowed not only to restrict science to the exclusive reference of the (actual) {\it res extensa}, it also allowed the introduction of the subject, the I, as a new realm characterized by the (potential) {\it res cogitans}. It was in Newton's physics that the potential mode of existence was completely erased from physical representation, producing an essential step in sedimenting this new duality between a material (physical) world and a conscious (human) world. In short, classical mechanics described nature as an {\it Actual State of Affairs}; i.e., a set of systems constituted by definite valued properties.\footnote{As discussed in detail in \cite{deRondeMassri18a}, an {\it Actual State of Affairs} (ASA) can be defined as a system constituted by a set of actual (definite valued) properties. Formally, a map from the set of properties to the $\{0,1\}$. Specifically, an ASA is a function $\Psi: \mathcal{G}\rightarrow\{0,1\}$ from the set of properties to $\{0,1\}$ satisfying certain compatibility conditions. Where the property $P\in \mathcal{G}$ is \emph{true} if $\Psi(P)=1$ and  $P\in \mathcal{G}$ is \emph{false} if $\Psi(P)=0$. The evolution of an ASA is formalized by the fact that the morphism $f$  satisfies $\Phi f=\Psi$.} This new picture of (physical) reality would be famously portrayed by the French mathematician Pierre-Simon Laplace: 
\begin{quotation} 
\noindent {\small ``Given for one instant an intelligence which could comprehend all the forces by which nature is animated and the respective situation of the beings who compose it ---an intelligence sufficiently vast to submit these data to analysis--- it would embrace in the same formula the movements of the greatest bodies of the universe and those of the lightest atom; for it, nothing would be uncertain and the future, as the past, would be present to its eyes. The human mind offers, in the perfection which it has been able to give to astronomy, a feeble idea of this intelligence. Its discoveries in mechanics and geometry, added to that of universal gravity, have enabled it to comprehend in the same analytical expressions the past and future states of the system of the world.'' \cite[p. 4]{Laplace02}} 
\end{quotation} 

Going back to the metaphysical premises of Greek atomism, Newton conceived the universe as constituted by many systems formed by smaller bodies (i.e., atoms) interacting within space and time. But now, through infinitesimal calculus and its rigorous mathematical definition of the notion of {\it continuum} it was possible to provide a new understanding of both space and time as the ``containers'' of all systems constituting the actual ``material'' world. But this was not the only unification that classical mechanics made possible. Far more important ---at least, from the perspective of the Greek tradition--- was the abstract unification between different perspectives or viewpoints. For the Greeks, the problem of motion, or how to characterize what is {\it the same} within {\it change} and {\it becoming}, was one of the kernel questions that attracted the interest of everyone searching for philosophical truth. If, as Heraclitus would argue, everything is in constant change and motion, if everything is always different to itself, how can we grasp any knowledge about anything? It was Aristotle who was able to create through the notion of {\it entity} as constituted by the principles of {\it existence}, {\it non-contradiction} and {\it identity}, a formal-conceptual representation in which knowledge about `things' would became possible. Aristotle's definition imposed a great level of metaphysical (and logical) abstraction, it talked about the existence, the non-contradiction and the identity of systems through their own change and motion. Now, in the 17th Century, through a precise definition of the notion of {\it reference frame}, it was also possible to produce a rigorous account of {\it invariance}. This implied the consistent unification of the One, {\it the same} state of affairs, as mathematically represented by the Many, the {\it different} reference frames. 

Invariance is an essential notion of physics allowing the production of a global representation of a state of affairs through the consistent translation between the different reference frames (or viewpoints) from which it is actually described. In this respect, it is not only the {\it invariant} static values of properties (e.g., mass, charge, etc.) but also of the {\it invariant variations} of dynamical properties (e.g., position, momentum, etc.) possessing distinct values in different {\it reference frames} which require a consistent translation between the multiple descriptions. Newtonian mechanics was able to account for such a consistency through a rigorous definition of the already existent ---so called--- Galilean transformations.\footnote{In the case of special relativity the translation between different reference frames is provided by the Lorentz transformations.} These transformations made possible the invariant type representation of the state of systems in such a way that all reference-frame dependent representations could be regarded as {\it different} and yet referring to the {\it the same} state of the object. In turn, this kernel idea would be also developed in conceptual terms. After David Hume's critical assessment of the notion of causality science was deathly wounded. Hume had recognized that empirical observations were incapable to ground a secure path for scientific knowledge. Causality was not something to be found within experience, it was something that we humans imposed to phenomena, the result of mental habit and custom. Thus, physical knowledge could be seen as derived from our ungrounded faith in induction and causation. Recognizing this relative aspect of representation, the physicist and philosopher Immanuel Kant would develop the notion of {\it objectivity} in order to restore ---at least partly--- the relevance of physical knowledge. Objectivity would secure scientific knowledge through the acceptance of a single common {\it representation} shared by all humans beings. This meant the acceptance of a {\it transcendental subject} shaped by a specific set of {\it categories} (grounded in Aristotelian metaphysics) and {\it forms of intuition} (Newton's space and time). But as a consequence, scientific knowledge, which since the Greeks had been always founded on the reference to {\it physis} (i.e., to reality), was now detached from reality and limited to the subject's capacity to account for his own experience about {\it objects}. Reality was re-framed as a the {\it Thing-in-Itself} (in German, {\it das Ding an sich}), something which physics was now incapable to grasp, and ``physical reality'' restricted to the subject's capacity to account for objective phenomena. According to Kant, the recognition that we humans are finite, should restrict the physicist to the acquisition of {\it objective knowledge}, namely, the knowledge coming from objects of experience. Kant had created a completely new architectonic for scientific knowledge in which reality could not be reached. This separation between knowledge and reality was the final conclusion of the Modern re-foundation of physics. A result that would certainly determine the path of Western thought in the centuries to come.



\section{The 20th Century Postmodern Re-Foundation of Physics}

Since the Enlightenment, the realist understanding of science would remain in constant siege. In Kant's co-relational metaphysics, reality was finally detached from scientific knowledge and replaced by the subject's capacity to account for objects of experience. Realism had been deathly wounded. Kant had argued that the sum of all objects, the empirical world, is a complex of appearances whose existence and connection occur only in our representations. Reality, renamed as a beast, {\it das Ding an sich} (the Thing-in-Itself), had survived but only as a monstrous paradoxical creature hiding beyond empirical sensibility, impossible to be known. As Kant would write in the {\it Prolegomena to Any Future Metaphysics}: ``And we indeed, rightly considering objects of sense as mere appearances, confess thereby that they are based upon a {\it thing-in-itself}, though we know not this thing as it is in itself, but only know its appearances, viz., the way in which our senses are affected by this unknown something.'' Kant had introduced the un-knowable within his own metaphysical system. Arthur Schopenahuer would make clear that the category of {\it causality} could not be applied within Kant's system to {\it noumenic reality}, which remained disconnected from categorical representation and (objective) experience. Friedrich Jacobi [1787: 223] famous remark would expose the problem in all its depth: ``Without the presupposition [of the `thing in itself,'] I was unable to enter into [Kant's] system, but with it I was unable to stay within it.'' Kant had introduced an essential separation between reality and theoretical representation. But it was still too soon for anti-realism to claim victory. Anti-realists would still have to wait two more centuries in order to rise as the supreme indisputable power. It is in our postmodern age, during the 20th Century, that realism would be finally defeated. And the main field of this final battle ---between realists and anti-realists--- would be no other than a new physical theory called quantum mechanics. 

The story of the second re-foundation of physics begins with the critical attack by the Austrian physicist and philosopher Ernst Mach to the inconclusive results of the 17th Century Modern re-foundation of physics. As described by Bas van Fraassen \cite[p. 2]{VF02}:  ``Kant exposed the illusions of Reason, the way in which reason overreaches itself in traditional metaphysics, and the limits of what can be achieved within the limits of reason alone. But on one hand Kant's arguments were not faultless, and on the other there was a positive part to Kant's project that, in his successors, engaged a new metaphysics. About a century later the widespread rebellions against the Idealist tradition expressed the complaint that Reason had returned to its cherished Illusions, if perhaps in different ways.'' It was by the end of the 19th Century that Mach would be able to produce the most radical deconstruction of the notions which grounded both Newton's physics and Kant's metaphysics. Through a return to empiricism, Mach developed a new positive scheme for physics which ---grounding itself on observability alone--- would attempt to completely erase any metaphysical remainings from the past. This meant the abandonment not only of Newtonian {\it absolute} space and time ---which acted as {\it a priori} notions in the Kantian architectonic--- but also the constitutive elements of the picture itself, namely, atoms. Mach was breaking the walls of the Modern spatiotemporal cage in which physics had been confined since Newton. But for him, the destruction of this jail implied also the destruction of metaphysics itself. Concomitant with Mach's deconstruction of classical physics, Friedrich Nietzsche would do the same for philosophy. Their common enemy had the same name: metaphysics. The success of both Mach and Nietzsche would leave {\it nothingness} as the new postmodern foundation. As a consequence, the 20th Century would be born in the turmoil of a deep crisis. Wolfgang Ernst Pauli, the godson of Ernst Mach, would reflect about his own time in the following manner: 
\begin{quotation} 
\noindent {\small ``In many respects the present appears as a time of insecurity of the fundamentals, of shaky foundations. Even the development of the exact sciences has not entirely escaped this mood of insecurity, as appears, for instance, in the phrases `crisis in the foundations' in mathematics, or `revolution in our picture of the universe' in physics. Indeed many concepts apparently derived directly from intuitive forms borrowed from sense-perceptions, formerly taken as matters of course or trivial or directly obvious, appear to the modern physicist to be of limited applicability. The modern physicist regards with scepticism philosophical systems which, while imagining that they have definitively recognised the {\it a priori} conditions of human understanding itself, have in fact succeeded only in setting up the {\it a priori} conditions of the systems of mathematics and the exact sciences of a particular epoch.'' \cite[p. 95]{Pauli94}} 
\end{quotation} 

The crisis produced by Mach's subversive deconstruction of classical mechanics had broken physics from its space-time chains. And it was certainly this liberation which would become essential for the development of both QM and relativity theory. While Albert Einstein applied Mach's positivist ideas in order to critically address the definition of {\it simultaneity} in classical mechanics, both Max Planck and Werner Heisenberg were able to advance new non-classical mathematical postulates and formalisms which could explicitly escape ---thanks to Mach's work--- the classical space-time representation of physics. In this context, a reloaded form of positivism, congregated in what was called the {\it Vienna Circle}, attempted to reshape Mach's main ideas and take them even further into the new century. The standpoint of their debates and discussions was of course grounded on Mach's own scheme:  
\begin{enumerate}
{\bf  \item[I.] Naive Empiricism:} Observation is a self evident {\it given} of ``common sense'' (macroscopic) experience. 
{\bf \item[II.] Physics as an Economy of Experience:} Physical theories are algorithmic mathematical formalisms which, through the addition of rules, produce predictions about observable measurement outcomes. 
{\bf \item[III.] Anti-Metaphysics:} Metaphysics, understood as an {\it interpretation} (or narrative) about unobservable entities, is not essentially required within empirically adequate theories. 
{\bf \item[IV.] Anti-Foundationalism:} Physics does not talk about an ``external reality'' beyond empirical observation. 
\end{enumerate}

The new quantum postulate proposed by Max Planck together with Einstein's special relativity gave a new force to the positivist agenda. After all, both theories had been developed following Mach's influential ideas. The Danish physicist Niels Bohr, maybe the most influential figure of the quantum revolution, would also play an essential role in this drama. Positivists together with Bohr would embrace the new postmodern {\it Zeitgeist} in which the possibility of a foundation for knowledge already seemed to everyone like a myth of the past. As Schr\"odinger world remark: ``The modern development [relativity and quantum mechanics], which those who have brought it to the fore are yet far from really understanding, has intruded into the relatively simple scheme of physics which towards the end of the nineteenth century looked fairly stabilized. This intrusion has, in a way, overthrown what had been built on the foundations laid in the seventeenth century, mainly by Galileo, Huygens and Newton. The very foundations were shaken''. In this context, the construction of the new postmodern scheme implied the essential addition of some extra points ---that went explicitly beyond Mach's scheme--- which both logical positivists and Bohr would carefully develop: 
\begin{enumerate}
{\bf  \item[V.] Intersubjective Justification:} The communication between the members of a scientific community allows to reach a consensus regarding the ``objective'' findings of a theory. 
{\bf \item[VI.] Atomist Metaphysics:} Metaphysical atomism is presupposed in anti-realist narratives as the microscopic (metaphysical) justification of our ``common sense'' macroscopic (empirical) experience. 
{\bf \item[VII.] Foundational Fictionalism:} Physics talks about an ``external reality'' beyond observability through the addition of {\it interpretations} (or narratives) to empirically adequate theories. 
\end{enumerate}

The attentive reader might have already recognized the various inconsistencies present within this scheme when considered as a whole. For example, {\bf III} and {\bf VI} are obviously in contradiction. While {\bf III} refuses to talk about reality beyond observability, {\bf V} makes reference to an atomist metaphysical picture in which unobservable particles are responsible not only for our `manifest image of the world' but also for the `clicks' we observe in detectors. The same happens with  {\bf IV} and {\bf VII}. While {\bf IV} argues against the possibility of a realistic foundation for science, {\bf VII} accepts reality as an essential reference for scientific discourse. These inconsistent aspects might be regarded at first sight as a flagrant weaknesses of the whole anti-realist program. However, strange as it might seem, exactly the opposite is true. Inconsistency is, in fact, the major strength of this postmodern scheme for it has allowed not only to include realism within a more general anti-realist program, but also to produce an incredibly efficient form of immunity against critical (realist) thought. Following {\it Michael Corleone's strategy:} ``Keep your friends close, and your enemies closer'', inconsistency has allowed to retain both realism and metaphysics within an explicitly anti-realist and anti-metaphysical system (see \cite{deRonde20c}). Both realists and metaphysicians have been forced to enter the gates of the new anti-realist architectonic completely unarmed. They have been welcomed to re-produce tasks which not only reinforce degraded, weak forms of realism, but also ---more importantly--- keep them away from any attempt of revolt. While metaphysicians have been confined to small prison cells and reassigned the role of `creating interpretations', realists who fanatically believe the newly created stories are left free to wonder in a maze with no exit, preaching these same naive fictions to the few fellow companions willing to listen. Anti-realists have even encouraged their captives to create more and more of these fantastic stories. Not only they enjoy watching the battles between realists, they are also certain that the power of their enemies is weakened with every new interpretation added to the maze. In a methodological level, inconsistency makes the whole program immune to scientific criticism for any critical remark can be easily lost when entering this labyrinth of contradicting paradoxical statements and undefined notions. Through the subversion of the main concepts of the realist scheme (i.e., reality, metaphysics, truth, objectivity, phenomena, measurement, etc.) any consistent attack is easily translated into a series of inconsistent statements, and then, immediately reintroduced as part of the anti-realist system itself. While at the theoretical level, inconsistency has become to be considered even as a value for science (see also \cite{BuenoVickers14, Meheus02, Vickers13}), in the practical individual level of research, inconsistency is accepted as a reasonable way of behaving for any member of the philosophical or scientific community. Everyone is free to argue in favor of {\bf A}, and immediately after change its mind in favor of {\bf not A}. While perspectivalism is encouraged, systematic theoretical consistency is regarded either as a kind of ``weakness'' or as a ``totalitarian'' attempt coming from fanatic realists who believe in a single truth. But while perspectivalism allows individuals to change continuously; the problems debated and their underlying presuppositions remain completely static. In this way, while heated debates take place in the safe un-grounded realm of ``beliefs'', ``interpretations'' and ``narratives'', the naive basis of ``empirical science'' remains unquestioned. 

The elasticity of postmodern philosophers and scientists allows them to continuously escape any solid statement. Apart from observations there is no true foundation to be found anywhere. As Popper puts it:
\begin{quotation}
\noindent {\small ``The empirical basis of objective science has thus nothing `absolute' about it. Science does not rest upon solid bedrock. The bold structure of its theories rises, as it were, above a swamp. It is like a building erected on piles. The piles are driven down from above into the swamp, but not down to any natural or `given' base; and if we stop driving the piles deeper, it is not because we have reached firm ground. We simply stop when we are satisfied that the piles are firm enough to carry the structure, at least for the time being.'' \cite[p. 111]{Popper92}} 
\end{quotation}
Moving freely in the swamp, anyone can choose the stance of their preference, change from one week to the next, or even deny that what was said was actually relevant. There is no system to be found. In this context, the destruction of theoretical unity has been accomplished through the replacement of theories by not necessarily consistent sets of fragmented models. As Nancy Cartwright \cite[p. 11]{Cartwright83} explains: ``In physics it is usual to give alternative theoretical treatments of the same phenomenon. We construct different models for different purposes, with different equations to describe them. Which is the right model, which the `true' set of equations? The question is a mistake. One model brings out some aspects of the phenomenon; a different model brings out others. Some equations give a rougher estimate for a quantity of interest, but are easier to solve. No single model serves all purposes best.'' The model-theoretic approach has provided the required logical framework in order to sustain the idea that theories are just ``sets of models'' \cite{daCostaFrench90}. An even more explicit erasure of the Greek notion of {\it theory} is also part of the contemporary research \cite{Vickers14}. 

As a result, physicists do not respect philosophers and philosophers do not take seriously the philosophical capacity of scientists. In this fragmented map the realist has no enemy, there are no referents to be found but many individual researches claiming opposite things. De-centralized, anti-realism appears as a ghost of many fragmented contradictory faces which can never be captured in a single standpoint. In the era of post-truth, where there are no secure foundations left to make any true statement, the anti-realist {\it status quo} is safer than ever. From a pedestal created by themselves, as ``specialists'' of a particular sub-sub-discipline that no one can question, the philosopher and the scientist explain to the layman: ``Those are my principles, and if you don't like them... well, I have others.'' 



\section{Re-Writing the History of Western Science}

In April 1966 Richard Feynman delivered an address to the {\it National Teachers' Association} in which he gave his fellow teachers lessons on how to teach their students to think like a scientist. Feynman used this opportunity to recall in a quite poetic manner the way in which his father ---a uniform salesman--- taught him the meaning of science by showing him how to derive conclusions from observations: 
\begin{quotation}
\noindent {\small ``[...] the result of observation, even if I were unable to come to the ultimate conclusion, was a wonderful piece of gold, with a marvelous result. It was something marvelous. [...] I think it is very important ---at least it was to me--- that if you are going to teach people to make observations, you should show that something wonderful can come from them. I learned then what science was about. It was patience. If you looked, and you watched, and you paid attention, you got a great reward from it (although possibly not every time). As a result, when I became a more mature man, I would painstakingly, hour after hour, for years, work on problems ---some times many years, sometimes shorter times--- many of them failing, lots of stuff going into the wastebasket; but every once in a while there was the gold of a new understanding that I had learned to expect when I was a kid, the result of observation. For I did not learn that observation was not worth while.'' \cite[p. 182]{Feynman99}} 
\end{quotation}
Today, the idea that theories are derived from observations is one of the very few generalized consensus that has been reached between scientists and philosophers. The widespread naturalization of this idea is certainly the most important accomplishment of the anti-realist re-foundation of science. An achievement which exposes the indisputable triumph of anti-realists over realists. The famous U.S. philosopher of physics Tim Maudlin, explains the same idea in a more sophisticated fashion: 
\begin{quotation} 
\noindent {\small ``Any empirical  science has to start from what the philosopher Wilfred Sellars called `the manifest image of the world'; that is, the world as it presents to me when I open my eyes. And of course we know that some of those appearances can be deceptive or misleading ---you know, a straw in water looks bent but it really isn't and so on ...--- but you have nowhere else to begin but with the manifest image, and then you try and produce theories that would explain it or account for it.'' \cite{Maudlin19b}} 
\end{quotation}

The uncritical acceptance of this naive empirical standpoint was introduced within science through a series of procedures involving not only the replacement of its reference, the carefully subversion of its main concepts but also the displacement of its historical origin. History is written by the victors, and the history of western science is no exception. By re-placing the origin of science from the 6th Century B.C. to the 16th Century A.D., anti-realists have been able not only to cut the ties of realists with their history, but also ---more importantly--- to re-write the history of science in their own terms as the history of ``empirical'' or ``modern'' science. Paying attention to these facts, one can then begin to understand the reason for the anti-realist aversion to the ancient Greek tradition as rooted in its undeniable reference to {\it physis} ---or, as we translate it today, reality. But also, maybe more importantly, the history of ancient Greek physics and philosophy tells us a story of triumphs of realists over sophists. Taking this into consideration, it is not strange to discover within our contemporary post-modern {\it Zietgeist} the constant attempt of anti-realists to try to forget a past of defeats. In this respect, it is interesting to notice, as Schr\"odinger would emphasize, that it was the father of the second anti-realist re-foundation of physics, Ernst Mach himself, who begun this offensive against the realist Greek tradition: 
\begin{quotation} 
\noindent {\small ``Ernst Mach, the physicist colleague of Gomperz at the University of Vienna, and eminent historian (!) of physics, had, a few decades earlier, spoken of the {\it `scarce and poor remnants of ancient science'.} He continues thus: {\it `Our culture has gradually acquired full independence, soaring far above that of antiquity. It is following an entirely new trend. It centres around mathematical and scientific enlightenment. The traces of ancient ideas, still lingering in philosophy, jurisprudence, art and science constitute impediments rather than assets, and will come to be untenable in the long run in face of the development of our own views'.} [...] In other passages of the same paper he recommends  quaint method of getting beyond antiquity, namely, to neglect and ignore it. In this, for all I know, he had little success ---fortunately, for the mistakes of the great, promulgated along with the discoveries of their genius, are apt to work serious havoc.'' \cite[pp. 20-21]{Schr54}} 
\end{quotation}
As we all know, Schr\"odinger derived his conclusion far too soon.

During the second half of the 20th century, specially with the rise of instrumentalism Mach's narrative would expand itself in order to become the orthodox view of science. An essential strategy for imposing this narrative was to attack the Middle Ages, presenting it as a dark irrational period in which religion and fantasy ruled the minds of disturbed people. Supplemented by a strong idea of progress and evolution the criticism to the Middle Ages was also ---implicitly--- pointing to everything that came before this period ---including the Greeks. As Feynman \cite[pp. 205-206]{Feynman99} himself would explain: ``During the Middle Ages there were all kinds of crazy ideas, such as that a piece of rhinoceros horn would increase potency. (Another crazy idea o f the Middle Ages is these hats we have on today-which is too loose in my case.) Then a method was discovered for separating the ideas-which was to try one to see if it worked, and if it didn't work, to eliminate it. This method became organized, of course, into science. And it developed very well, so that we are now in the scientific age.'' In this new empiricist narrative, special emphasis has been given to the figures of Newton and Galileo as the heroes who finally ended the dark ruling of Aristotelian metaphysics. By simply making no reference to neo-Platonism its undeniable historical influence was simply erased. Just to give one example between many, the world famous English physicist Stephen Hawking writes in his best-seller, {\it A Brief History of Time}:
 \begin{quotation} 
\noindent {\small ``Our present ideas about the motion of bodies date back to Galileo and Newton. Before them people believed Aristotle, who said that the natural state of a body was to be at rest and that it moved only if driven by a force or impulse. It followed that a heavy body should fall faster than a light one, because it would have a greater pull toward the earth. The Aristotelian tradition also held that one could work out all the laws that govern the universe by pure thought: it was not necessary to check by observation. So no one until Galileo bothered to see whether bodies  of different weight did in fact fall at different speeds.'' \cite[p. 15]{Hawking88}} 
\end{quotation} 
The effect of this passage is twofold. Firstly, it places the beginning of our own scientific tradition in the 16th Century A.D. erasing the importance of the Ancient Greeks. Secondly, it presents Aristotle as someone who was not capable of testing experimentally his own  ---quite naive--- theory of motion, exposing in this way the weakness of a supposedly more primitive tradition. This dismissive reference to the Greeks is completely widespread in the present scientific and philosophical literature where the ancient philosophers are depicted as presenting very primitive ideas, stressing in this way our own contemporary grandiosity. The U.S. physicist Michio Kaku provides an even more explicit presentation of the orthodox narrative: 
\begin{quotation} 
\noindent {\small ``The history of physics is the history of modern civilization. Before Isaac Newton, before Galileo, we were shrouded with the mysteries of superstition. People believed in all sorts of different kinds of spirits and demons. What made the planets move? Why do things interact with other things? It was a mystery. So back in the Middle Ages, for example, people read the works of Aristotle. And  Aristotle asked the question, `Why do objects move toward the earth? And that's because,' he said, `objects yearn, yearn to be united with the earth. And why do objects slow down when you put them in motion? Objects in motion slow down because they get tired.' These are the works of Aristotle, which held sway for almost 2000 years until the beginning of modern physics with Galileo and Isaac Newton.'' \cite{Kaku12}} 
\end{quotation} 
In the same line, Steven Weinberg \cite{Weinberg88} remarks in an interview from 1988 the essential discontinuity between the Moderns and everything that came before: ``I think that science has changed the way that we think about ourselves, about our role in the universe. I think there's lots of historical evidence for the tremendous change in Western society that is caused by the scientific revolution of the 16th and 17th century. You know, we stopped burning witches. I think the changes are incalculable.''  Almost three decades later, in {\it To Explain the World: The Discovery of Modern Science}, he also points out that ``neither the Milesians nor any of the other Greek students of nature were reasoning in anything like the way scientists reason today.'' In the {\it Preface} of his book Weinberg remarks:
\begin{quotation} 
\noindent {\small ``Before history there was science, of a sort. At any moment nature presents us with a variety of puzzling phenomena: fire, thunderstorms, plagues, planetary motion, light, tides, and so on. Observation of the world led to useful generalizations: fires are hot; thunder presages rain; tides are highest when the Moon is full or new, and so on. These became part of the common sense of mankind. But here and there, some people wanted more than just a collection of facts. They wanted to explain the world. It was not easy. It is not only that our predecessors [before Modern Science] did not know what we know about the world ---more importantly, they did not have anything like our ideas of what there was to know about the world, and how to learn it.'' \cite{Weinberg15}} 
\end{quotation} 


Of course, this particular re-writing of the history of science is not only essential in oder to erase the kernel link of between science and reality, it is also essential in order to create a narrative in which theories begin with observations and end with their prediction. As described by Hawking: 
\begin{quotation} 
\noindent {\small ``In order to talk about the nature of the universe and to discuss questions such as whether it has a beginning or an end, you have to be clear about what a scientific theory is. I shall take the simpleminded view that a theory is just a model of the universe, or a restricted part of it, and a set of rules that relate quantities in the model to observations that we make. It exists only in our minds and does not have any other reality (whatever that might mean). A theory is a good theory if it satisfies two requirements. It must accurately describe a large class of observations on the basis of a model that contains only a few arbitrary elements, and it must make definite predictions about the results of future observations. For example, Aristotle believed Empedocles's theory that everything was made out of four elements, earth, air, fire, and water. This was simple enough, but did not make any definite predictions. On the other hand, Newton's theory of gravity was based on an even simpler model, in which bodies attracted each other with a force that was proportional to a quantity called their mass and inversely proportional to the square of the distance between them. Yet it predicts the motions of the sun, the moon, and the planets to a high degree of accuracy.'' \cite[p. 10]{Hawking88}} 
\end{quotation} 

Placing themselves as the evolved figures of true Modern progress, in a tradition that begins with the Enlightenment, contemporary anti-realist scientists and philosophers of science have attempted systematically to escape any direct reference to the ancient Greek tradition. According to this narcissistic narrative, we ourselves have more knowledge and understanding than any other community in the history of human kind. We are the proud result of a progressive evolution that begun in 16th Century and has lead us to the peak of empirical scientific understanding and knowledge. The justification of what is considered to be a ``self evident'' fact is twofold. From a quantitative perspective, the great specialization within science shows that we have ``more knowledge'' than ever before. So we are told, we have ``so much knowledge'' that, unlike in the past, maybe for the last time with Leonardo and Leibniz, a single human cannot grasp all the scientific knowledge that we have conquered. Today, it is accepted that one cannot even gain the complete knowledge of a single discipline, like physics. From a pragmatic perspective, today's fantastic technological capacity, which even allows us to destroy the whole world by pressing a single button, exposes the true power of our knowledge. Never before have we been able to compute the quantity of information at the speed and efficiency we can do today. Both ideas are of course grounded in a deep confusion between technological power and scientific understanding, between quantity and quality. While technology implies a  `know how', a practical capacity, it does not provide a representation of reality. While models might be able to provide an operational solution to very specific problems, they are certainly unable to provide scientific understanding through consistent, coherent and unified theoretical representations. 

As it is well known, most of the founding fathers of QM ---exception made of Paul Dirac--- would have never endorsed this anti-realist narrative of the history of science. Most of them had been trained in the European {\it Gymnasium} which imposed a deep knowledge of the Greek tradition, not only restricted to its language but also to its cultural heritage. Heisenberg, Pauli and Schr\"odinger ---the major figures of the quantum revolution--- constantly stressed the essential reference of their own developments in physics to those of the first philosophers (see for a detailed analysis \cite{FM20}). Not without a reason, many of them have a book called {\it Physics and Philosophy}. Werner Heisenberg \cite[p. 57]{Heis58b} would argue that: ``I saw the achievements of modern times, of Newton and his successors, as the immediate consequence of the efforts of the Greek mathematicians and philosophers, and never once did it occur to me to consider the science and technology of our times as belonging to a world basically different from that of the philosophy of Pythagoras and Euclid.'' Heisenberg would go back to the Greeks constantly not only as intellectual adventure, but more importantly to understand the theory of quanta itself, for he believed that ``one could hardly make progress in modern atomic physics without a knowledge of Greek natural philosophy.'' Wolfgang Pauli \cite[p. 40]{Pauli94} would also constantly refer to the Greeks: ``The critical scientific spirit however reached its first culmination in classical Hellas. It was there that those contrasts and paradoxes were formulated which also concern us as problems, though in altered form: appearance and reality, being and becoming, the one and the many, sense experience and pure thought, the continuum and the integer, the rational ratio and the irrational number, necessity and purposefulness, causality and chance'' And the same with Erwin Schr\"odinger \cite[pp. 17-18]{Schr54} who would stress that: ``the thinkers who started to mould modern science did not begin from scratch. Though they had little to borrow from the earlier centuries of our era, they very truly revived and continued ancient science and philosophy.'' Seeking for help, he would ---just like Heisenberg and Pauli--- return to the Greeks: ``By the serious attempt to put ourselves back into the intellectual situation of the ancient thinkers [...] we may regain from them their freedom of thought.'' We might address a passage from Raimundo Fern\'andez Mouj\'an which clearly addresses the issue at stake: 
\begin{quotation} 
\noindent {\small ``As we revise their quotes, we are made aware of the striking contrast between the views of Schr\"odinger, Pauli and Heisenberg, and the widely accepted vision today in scientific contexts, according to which science, properly speaking, originated in modern times, in the XVIIth century. These physicists from the early XXth century did not see what began in the XVIIth century and was developed over the following centuries as the origin of science, but rather as the origin of the specific parameters of modern science that quantum mechanics, among other developments, was rejecting. Illuminism, however admirable, was for them the establishment of a certain view of science that had become problematic, and they saw better allies for their task among the ancient Greeks. They found Plato, Pythagoras or Heraclitus more useful than the modern philosophers and physicists. They were insistent critics of the parameters of science established in the XVIIth, XVIIIth and XIXth centuries. In their works they frequently make this contrast explicit, they distinguish between the aspects of science that had their origin in the XVIIth century and that seemed now problematic (the separation of the world in res extensa and res cogitans, the limiting of natural science to a narrow materialism, the sharp separations among disciplines and faculties, etc.) and the ancient Greek model, which could help them overcome those obstacles.'' \cite{FM20}} 
\end{quotation}

Last but not least, Albert Einstein can be also linked to the Greek tradition maybe not directly but certainly through his admiration for Baruch Spinoza who stood close to the unity, consistency and coherency of physical representation, fighting ---also, in almost complete solitude--- the Modern attack to the scientific knowledge about reality ---or, in Spinozian terms, about Nature. In this respect, Spinoza can be considered as a Modern philosopher who ---rejecting the ontological division between {\it res cogitans} and {\it res extensa}--- stood close to the ancient Greek tradition of the One (i.e., the only one substance) and the Many (i.e., the infinitely many attributes, namely, what the intellect perceives of substance as constituting its essence). As Einstein \cite[p. 373]{Viereck} would explain: ``I am fascinated by Spinoza's Pantheism. I admire even more his contributions to modern thought. Spinoza is the greatest of modern philosophers because he is the first philosopher who deals with the soul and the body as one, not as two separate things.''

\section{The Postmodern Physical Attack to Philosophy}

Just like sophists ---in the 4th Century B.C.---, contemporary anti-realists ---since the second half of the 20th Century--- having already conquered not only physics but the totality of science, engaged in a furious attack against an already defeated philosophy. The reason might be found in the serious threat that philosophy would represent for the anti-realists establishment of ``empirical science'' if awaken from its comatose state. For it is only philosophy which can expose the dogmatic-naive basis of the present anti-realist understanding of science. It is the conceptual (or metaphysical) level of scientific observation which anti-realists ---just like with the historical reference to the ancient Greeks--- constantly forget to address. As Einstein explains:
\begin{quotation}
\noindent {\small ``From Hume Kant had learned that there are concepts (as, for example, that of causal connection), which play a dominating role in our thinking, and which, nevertheless, can not be deduced by means of a logical process from the empirically given (a fact which several empiricists recognize, it is true, but seem always again to forget). What justifies the use of such concepts? Suppose he had replied in this sense: Thinking is necessary in order to understand the empirically given, {\it and concepts and `categories' are necessary as indispensable elements of thinking.}'' \cite[p. 678]{Einstein65} (emphasis in the original)}
\end{quotation} 
Exactly this same point was repeatedly stressed by Heisenberg \cite[p. 264]{Heis73} in several of his elder writings during the 1970s: ``The history of physics is not only a sequence of experimental discoveries and observations, followed by their mathematical description; it is also a history of concepts. For an understanding of the phenomena the first condition is the introduction of adequate concepts. Only with the help of correct concepts can we really know what has been observed.'' Wolfgang Pauli \cite[p 129]{Pauli94} would argue in a more sarcastic manner: ``I hope no one still maintains that theories are deduced by strict logical conclusions from laboratory-books, a view which was still quite fashionable in my student days.'' Confronting these ``philosophical ideas'' Richard Feynman would explicitly attack, during an interview he gave in 1979, the need of such conceptual (or metaphysical) requirements: ``[philosophers] seize on the possibility that there may not be any ultimate fundamental particle, and say that you should stop work and ponder with great profundity. `You haven't thought deeply enough, first let me define the world for you.' Well, I'm going to investigate it {\it without} defining it!'' Since then, Feynman's anti-philosophical viewpoint has become part of orthodoxy. Quite regardless of the forgotten remarks made by Einstein, Heisenberg, Pauli, Hanson and many others, the orthodox contemporary consideration of theories as derived from observation has become not only dogma but also the standpoint of all contemporary debates.\footnote{Maybe the only contemporary physicist who has dare to go against orthodoxy is David Deutsch \cite[pp. 3-4]{Deutsch04} who in his book, {\it The Beginning of Infinity}, writes:  ``The fact that the light was emitted very far away and long ago, and that much more was happening there than just the emission of light ---those are not things that we see. We know them only from theory. Scientific theories are explanations: assertions about what is out there and how it behaves. Where do these theories come from? For most of the history of science, it was mistakenly believed that we `derive' them from the evidence of our senses ---a philosophical doctrine known as empiricism.''} Many decades later, Feynman's student and biographer, the influential U.S. cosmologist and popular-science writer Lawrence Krauss \cite{Krauss12} is able to claim with great confidence: ``Physics is an empirical science. As a theoretical physicist I can tell you that I recognize that it's the experiment that drives the field, and it's very rare to have it go the other way; Einstein is of course the obvious exception, but even he was guided by observation. It's usually the universe that's surprising us, not the other way around.'' It is from this naive empirical standpoint which understands observation as unproblematic that, since the rise of instrumentalism,  contemporary scientists have continuously and vigorously attacked the ``utility'' of philosophy. 

Empirical scientists are essentially correct. Given you (mis)understand theories as mathematical schemes capable to predict observations, there is no role what so ever to be played by philosophy within science. The natural conclusion which Stephen Hawking \cite{Hawking11} has been honest enough to state clearly during a talk in 2011 is that: ``Philosophy is dead. Philosophers have not kept up with modern developments in science. Particularly physics.'' During the 1990s the U.S. Nobel laureate Steven Weinberg, wrote in his famous book, {\it Dreams of a Final Theory:}
\begin{quotation} 
\noindent {\small ``The insights of philosophers have occasionally benefited physicists, but generally in a negative fashion ---by protecting them from the preconceptions of other philosophers. I do not want to draw the lesson here that physics is best done without preconceptions. At any one moment there are so many things that might be done, so many accepted principles that might be challenged, that without some guidance from our preconceptions one could do nothing at all. It is just that philosophical principles have not generally provided us with the right preconceptions. [...] Physicists do of course carry around with them a working philosophy. For most of us, it is a rough-and-ready realism, a belief in the objective reality of the ingredients of our scientific theories. But this has been learned through the experience of scientific research and rarely from the teachings of philosophers. This is not to deny all value to philosophy, much of which has nothing to do with science. I do not even mean to deny all value to the philosophy of science, which at its best seems to me a pleasing gloss on the history and discoveries of science. But we should not expect it to provide today's scientists with any useful guidance about how to go about their work or about what they are likely to find.'' \cite{Weinberg93}} 
\end{quotation}
In an interview from 2004, Neil deGarsse Tysson, the famous U.S. astro-physicist and host for the new series, {\it Cosmos: A Spacetime Odyssey},\footnote{A follow-up to the 1980s TV series presented by Carl Sagan.} also addressed the futility of philosophy for science: 
\begin{quotation} 
\noindent {\small ``if you are distracted by your questions so that you can't move forward, you are not being a productive contributor to our understanding of the natural world. And so the scientist knows when the question `What is the sound of one hand clapping?' is a pointless delay in our progress. [Insert predictable joke by one interviewer, imitating the clapping of one hand.] How do you define `clapping'? All of a sudden it devolves into a discussion of the definition of words. And I'd rather keep the conversation about ideas. And when you do that, don't derail yourself on questions that you think are important because philosophy class tells you this. The scientist says [to the philosopher], `Look, I got all this world of unknown out there. I'm moving on. I'm leaving you behind. You can't even cross the street because you are distracted by what you are sure are deep questions you've asked yourself. I don't have the time for that'.'' \cite{Pigliucci14}} 
\end{quotation}
Continuing the attack, Lawrence Krauss has even argued that:  
\begin{quotation} 
\noindent {\small ``Philosophy is a field that, unfortunately, reminds me of that old Woody Allen joke, `those that can't do, teach, and those that can't teach, teach gym.' And the worst part of philosophy is the philosophy of science; the only people, as far as I can tell, that read work by philosophers of science are other philosophers of science. It has no impact on physics what so ever. ... they have every right to feel threatened, because science progresses and philosophy doesn't.'' \cite{Krauss12}} 
\end{quotation}

The truth is that apart from a very small number of scientists interested in philosophical issues ---often confused with the divulgation of science or with scientific explanations ``for dummies''---, the vast majority of contemporary scientists don't even bother to take seriously philosophical problems. They simply do not relate their work ---in any way--- with philosophy. Period. Unlike the founding fathers of QM, physicists today have no knowledge about philosophy what so ever and ---even worse--- they do not think it would be important for them to learn the philosophies of Kant, Spinoza, Plato or ---even--- Mach in order to better understand their own discipline. Once again, Richard Feynman is a great example of the widespread attitude of contemporary physicists towards philosophy:
\begin{quotation} 
\noindent {\small ``My son is taking a course in philosophy, and last night we were looking at something by Spinoza and there was the most childish reasoning! There were all these attributes, and Substances, and all this meaningless chewing around, and we started to laugh. Now how could we do that? Here's this great Dutch philosopher, and we're laughing at him. It's because there's no excuse for it! In the same period there was Newton, there was Harvey studying the circulation of the blood, there were people with methods of analysis by which progress was being made! You can take every one of Spinoza's propositions, and take the contrary propositions, and look at the world and you can't tell which is right. Sure, people were awed because he had the courage to take on these great questions, but it doesn't do any good to have the courage if you can't get anywhere with the question.''  \cite[p. 195]{Feynman99}} 
\end{quotation}
Steven Weinberg, certainly one of the most important and influential figures of contemporary physics, makes reference to Plato and Aristotle in the following manner: ``I confess that I find Aristotle frequently tedious, in a way that Plato is not, but although often wrong Aristotle is not silly, in the way that Plato sometimes is.'' According to the U.S. Nobel laurate and father of the {\it Standard Model}: ``none of the early Greeks from Thales to Plato, in either Miletus or Abdera or Elea or Athens, ever took it on themselves to explain in detail how their theories about ultimate reality accounted for the appearances of things. This was not just intellectual laziness. There was a strain of intellectual snobbery among the early Greeks that led them to regard an understanding of appearances as not worth having.'' As Weinberg explains, there is an essential difference between the Greeks and modern science:  
\begin{quotation} 
\noindent {\small ``There is an important feature of modern science that is almost completely missing in all the thinkers I have mentioned, from Thales to Plato: none of them attempted to verify or even (aside perhaps from Zeno) seriously to justify their speculations. In reading their writings, one continually wants to ask, ``How do you know?' [...] Aristotle called the earlier Greek philosophers physiologi, and this is sometimes translated as `physicists,' but that is misleading. The word physiologi simply means students of nature (physis), and the early Greeks had very little in common with today's physicists. Their theories had no bite. Empedocles could speculate about the elements, and Democritus about atoms, but their speculations led to no new information about nature ---and certainly to nothing that would allow their theories to be tested. It seems to me that to understand these early Greeks, it is better to think of them not as physicists or scientists or even philosophers, but as poets.'' \cite{Weinberg15}} 
\end{quotation}

It should come with no surprise that since the coming into power of instrumentalism after the war, professors in physics have systematically prohibited philosophical thinking of students within classrooms. Quite recently, not only philosophers of physics who studied physics like David Albert and Tim Maudlin but also physicists like Lee Smolin and Sean Carroll have come forward to tell their stories of intellectual oppression and harassment. In several interviews they have revealed something that every physicist already knows, namely, that in our times it has become dangerous for students in physics to engage in philosophical reflections about physics. As Smolin remembers:  
\begin{quotation} 
\noindent {\small ``When I learned physics in the 1970s, it was almost as if we were being taught to look down on people who thought about foundational problems. When we asked about the foundational issues in quantum theory, we were told that no one fully understood them but that concern with them was no longer part of science. The job was to take quantum mechanics as given and apply it to new problems. The spirit was pragmatic; `Shut up and calculate' was the mantra. People who couldn't let go of their misgivings over the meaning of quantum theory were regarded as losers who couldn't do the work.'' \cite[p. 312]{Smolin07}} 
\end{quotation}
In a recent interview by Sean Carroll, David Albert \cite{Albert19} has recalled a very disturbing anecdote which should be seriously addressed: 
\begin{dialogue}
 {\small 
\speak{Sean Carroll} {\it And I'm gonna re-write that anecdote, it's gonna be no good. And it's about your experience as a graduate student at Rockefeller and you made the mistake of reading a book by David Hume, I understand correctly. And that got you interested in foundations and what happened?}

\smallskip

\speak{David Albert} {\it Well, that wasn't received warmly at Rockefeller.} 

\smallskip

\speak{Sean Carroll} {\it This was in the Physics department?} 

\smallskip

\speak{David Albert} {\it This was in the Physics department at Rockefeller. There were within fairly short order proceedings instituted to expel me from the PhD program. This was back in the late, very late 1970s, early 1980s. So this was in dark old days, but worrying about issues like that was profoundly unpopular. I got to stay in the program on the condition that I would work on a thesis topic assigned to me by the department, instead of one that I chose, and the one that was assigned was clearly one that was thought to be good for my character.} 

\smallskip

\speak{Sean Carroll} {\it Well, it turned out okay. Your character is still good today.} 

\smallskip

\speak{David Albert} {\it It was an extremely calculation heavy, it was something about Borel resummation and flight of the fourth field theories, which people were playing around with in those days. It was an extremely computation-heavy thesis topic. I had been in contact at that time already with Yakir Aharonov, who I had written to while I was a graduate student, telling him some questions that I was struggling with and didn't know what to do with. I asked for his advice. He was wonderful, and I did go to him, I was given a pretty stark choice at a meeting in the Dean's office, that either I was gonna do a thesis topic assigned to me by the department or I was gonna leave the PhD program.}}
\end{dialogue}
It should be clear that the attempt to expel a student from a PhD program because he read a book should not be considered as part of the scientific enterprise. The scientific quest says nothing about ``correcting the character'' of young students eager to learn about the mysteries of nature. On the contrary, scientific training should encourage students to become revolutionaries, to think beyond what has been thought ever before. The situation that Albert describes is something quite common within physics departments and has not yet stooped. As Carroll \cite{Carroll20} has recently described: ``Many people are bothered when they are students and they first hear [about Standard Quantum Mechanics]. And when they ask questions they are told to shut up. And if they keep asking they are asked to leave the field of physics.'' I myself, as a student in physics during the late 1990s, sharing the same forbidden inclinations towards philosophical realist thought, experienced and witnessed the same intellectual harassment to students who would dare to ask forbidden questions or attempt to produce a more conceptual investigation than required. And, of course, I was later on also invited to leave the field of physics. 

It is these same group of bullied physicists, some of them transformed into philosophers ---I am myself one of them---, who have attempted to defend philosophy from their ---very well known--- aggressors. Unfortunately, their defense has avoided any direct confrontation with the contemporary anti-realist and anti-philosophical understanding of ``empirical science''. In fact, all these philosophers and physicists who attempt to defend philosophy take for granted that science and philosophy are completely different fields of research which do not intersect at any point. As explained by one of the defenders, the U.S. philosopher and former scientist Massimo Pigliucci \cite{Pigliucci09}: ``Science, broadly speaking, deals with the study and understanding of natural phenomena, and is concerned with empirically (i.e., either observationally or experimentally) testable hypotheses advanced to account for those phenomena. Philosophy, on the other hand, is much harder to define. Broadly speaking, it can be thought of as an activity that uses reason to explore issues that include {\it the nature of reality} (metaphysics), the structure of rational thinking (logic), [etc.]'' Sean Carroll \cite{Carroll14}, a physicist who actively supports philosophy of physics, explains: ``Nobody denies that the vast majority of physics gets by perfectly well without any input from philosophy at all. (`We need to calculate this loop integral! Quick, get me a philosopher!')'' Anyhow, according to some defenders the problem seems to be that scientists do not understand what is the role played by philosophy. In fact, this is exactly what Pigliucci \cite{Pigliucci12} recriminates to Lawrence Krauss: ``Krauss does not understand what the business of philosophy is.'' Addressing the attack, Wayne Myrvold, a philosopher of physics has recently ``asked a few physicists [who have a much more positive attitude towards philosophy] to write a few words about why they regard talking to philosophers valuable.'' Carlo Rovelli \cite{Rovelli14}, a famous Italian physicist with clear philosophical inclinations, accepting Myrvold's challenge has used this opportunity to reply to deGrasse: ``Neil deGrasse Tyson is not the only one to consider philosophy useless for science. Many of my colleagues in science have the same view, today. But they would be perhaps surprised to find out that many scientists have the opposite view. Among these are names like Einstein, Heisenberg, Bohr, Darwin and Newton, just to name a few.  Why do these major scientists, and so many others of lesser stature, consider and still consider philosophy as an extremely useful enterprise in our search for knowledge?  In my opinion, because they understood the nature of the scientific enterprise better than Tyson.'' Smolin is another physicist who tried to justify why it might be interesting to talk to talk to philosophers:
\begin{quotation} 
\noindent {\small ``The triumph of the pragmatic style in the late 1940s and 1950s marginalized the more foundational style, pushing its exponents out of the elite centres and into smaller universities.  The main venue for philosophical reflection in the theoretical physics community became the relativity community, which valued breadth and independence of thought, and which thrived through the 1980s in perhaps a dozen centres, until its foundational orientation was destroyed by the NSF's pragmatic push to fund support for LIGO and numerical relativity.  Meanwhile, particle physics developed rapidly in the pragmatic mode, with great triumphs until progress came to a halt in the early 1980s with the failed searches for proton decay and other signs of unification beyond the standard model. 

Since then, fundamental physics has been in a crisis, due to the evident need for new revolutionary ideas ---which becomes more evident with each failure of experiment to confirm fashionable theories, and the inability of those trained in a pragmatic, anti-philosophical style of research to free themselves from fashion and invent those new ideas. To aspire to be a revolutionary in physics, I would claim, it is helpful to make contact with the tradition of past revolutionaries. But the lessons of that tradition are maintained not in the communities of fashionable science, with their narrow education and outlook, but in the philosophical community and tradition.  And that is why I talk with philosophers and encourage my students to do so.'' \cite{Smolin14}} 
\end{quotation}
And yet, playing the devils advocate, one could argue that many scientists have also found inspiration in literature, music and many other fields rather than philosophy. So it might be also interesting to talk to musicians, artists and writers. So what is so special about philosophy? 

This question takes us back to the most obvious line of argumentation for a proper defense of philosophy, namely, to argue that there is in fact a deep relation between philosophy and science grounded on their common subject of inquiry: reality. This is why it is not strange to find out that the defense of philosophy has been undertaken only by realists. As Carroll \cite{Carroll14} describes: ``Many of the best philosophers of physics were trained as physicists, and eventually realized that the problems they cared most about weren't valued in physics departments, so they switched to philosophy. But those problems ---the basic nature of the ultimate architecture of reality at its deepest levels--- are just physics problems, really. And some amount of rigorous thought is necessary to make any progress on them. Shutting up and calculating isn't good enough.'' Smolin \cite[p. 7]{Smolin07} argues in a similar vein: ``Physics should be more than a set of formulas that predict what we will observe in an experiment; it should give a picture of what reality is.'' Rovelli \cite{Rovelli14} reinforces the point: ``if we understand science as a technical machine for collecting data and testing theories, then we do not need much philosophy.  But science is not just that.'' As he \cite{Rovelli07} explains: ``Science is a continuous quest for the best way to think about and look upon the world. It is, above all, an ongoing exploration of new forms of thinking.'' Tim Maudlin \cite{Maudlin19b} illustrates these same ideas in the context of QM: ``There is no doubt and no one is going to dispute that there's a mathematical formalism that we know how to derive predictions from, and those predictions can be accurate to 14 decimal places. But what a theory is, what a physical theory is, is more than just a mathematical formalism with some rules. It should specify a physical ontology which means: tell me what exists in the physical world. Are there particles? Are there fields? Is there space-time? And tell me about those things and then specify some laws about how they behave, that tell me how they behave through time.'' In order to address reality philosophers of physics have introduced {\it interpretations} within theories. Interpretations are stories added to empirically adequate theories which make explicitly clear what these theories are really talking about, to what entities they refer. As David Albert \cite{Albert12} explains, when attempting to develop these narratives ``the problems that we run into are peculiar philosophical sorts of problems. What is going to count as a solution to this problem? What's the minimum that we need in order to tell a story[?] [...] These are questions of epistemology, questions of our relationship, of our beliefs to the external world ---and so on and so forth--- that philosophers have been dealing with for a log time.'' So it might seem, we have finally found the missing link between empirical science and philosophical research. Well, not to fast... As Bas van Fraassen \cite[p. 10]{VF80}, one of the most prominent anti-realist contemporary figures, has warned (realist) philosophers and physicists, ``theories need not be true to be good. [...] To develop an empiricist account of science is to depict it as involving a search for truth only about the empirical world, about what is actual and observable.''  As he \cite[p. 242]{VF91} concludes: ``However we may answer these questions [about interpretation], believing in the theory being true or false is something of a different level.'' With a gentle reminder van Fraassen has easily collapsed the house of cards that realists were so carefully attempting to construct. Van Fraassen is absolutely right, the realist program has never been able to produce an objective set of rules that would allow anyone to choose ---beyond subjective preferences--- between the many interpretations, the one which is actually true (see for a detailed discussion \cite{deRonde20c}). This kernel problem is even acknowledged by Maudlin \cite{Maudlin19b} who recognizes that: ``different approaches would give you different answers [to the question of physical reality in QM]. My preference, my aesthetic sense is that the pilot wave seems more natural to me and the objective collapse seems to me more unnatural but I wouldn't ---you know---give a lot of attention to my aesthetic preferences here.'' Given that there is no way to choose an interpretation and that it is not even clear what would be the gain of adding these fictional stories, the conclusion reached by Arthur Fine \cite[p. 149]{Fine86} seems to us quite reasonable: ``Try to take science on its own terms, and try not to read things into science. If one adopts this attitude, then the global interpretations, the `isms' of scientific philosophies, appear as idle overlays to science: not necessary, not warranted and, in the end, probably not even intelligible.'' It is at this point, that realists run back to their pragmatic shelter from which they should have never attempt to escape and recognize, like Carroll \cite{Carroll12}, that: ``reality is a useful way of talking about the actual world'' and ---in the end--- ``our description of reality is contingent on utility''.

\section{The Dogmatic-Naive ``Understanding'' of Quantum Mechanics} 

The anti-realist re-foundation of physics that took place in the 20th Century finds its most clear exposition in the process of ``standardization'' of QM that took place during the second and third decades. The result of this operation ended in an inconsistent textbook formulation ---including collapses, measurements and outcomes--- which has been taught ever since in Universities all around the world. As Arthur Fine has described with a triumphant tone: 
\begin{quotation}
\noindent {\small ``These instrumentalist moves, away from a realist construal of the emerging quantum theory, were given particular force by Bohr's so-called `philosophy of complementarity'; and this
nonrealist position was consolidated at the time of the famous
Solvay conference, in October of 1927, and is firmly in place today.
Such quantum nonrealism is part of what every graduate physicist
learns and practices. It is the conceptual backdrop to all the
brilliant success in atomic, nuclear, and particle physics over the
past fifty years. Physicists have learned to think about their
theory in a highly nonrealist way, and doing just that has brought
about the most marvelous predictive success in the history of
science.'' \cite[p. 1195]{PS}}
\end{quotation}
Indeed, the Standard version of QM, sometimes also referred to by physicists as the ``Copenhagen interpretation'', was imposed by an anti-realist program through a series of ``instrumentalist moves'' which had the essential purpose to replace the need of a consistent representation by the reference to measurements, outcomes and ---even--- strange collapses. Bohr's introduction of his irrepresentable ``quantum jumps'' in his famous inconsistent model of the hydrogen atom, Born's probabilistic reference to measurement outcomes in Schr\"odinger's wave mechanics, Bohr's creation of the principles of {\it complementarity} and {\it correspondence} as well as their application to Heisenberg's renamed ``uncertainty'' relations are just some of these moves. We could also mention Dirac's and von Neumann's explicit introduction of the famous ``measurement rule'' within the axiomatic formulation of the theory which, in turn, lead most philosophical investigations into endless debates about the infamous ``measurement problem'' as well as Bohr's famous reply to the EPR paper which crowned him as the champion of the still ongoing anti-realist era (for a detailed analysis see \cite{deRonde20, deRonde20b}). However, as we have already stressed, the success of the anti-realist program is intrinsically related to its tolerance to the (realist) metaphysical discourse about a microscopic realm constituted by elementary particles. The uncritical acceptance that `clicks' in detectors are good enough evidence for the existence of `quantum particles' has played an essential role in the anti-realist strategy of control. It can hardly be overestimated that, as a matter of fact, the reference to a microscopic (quantum) realm as well as the `clicks' in detectors has become a contemporary dogma followed by both realists and anti-realists, physicists and philosophers. All interpretations of QM share exactly this same underlying atomist metaphysics supplemented by a naive understanding of experience. 

It is this completely unjustified but widespread idea according to which QM talks about measurement outcomes which are the consequence of our interaction with elementary particles which combines the worst of both realist and anti-realist worlds, namely, metaphysical dogmatism and naive empiricism. While metaphysical dogmatism imposes the idea that the theory must obviously talk about microscopic `particles', naive empiricism dictates that what we evidently observe in the lab are single measurement outcomes ---i.e., clicks' in detectors or `spots' in photographic plates instead of intensive patterns of spectrum lines (see for a detailed analysis \cite[Sect. 4]{deRonde20b}). This dogmatic-naive restriction to the definitions of both physical reality and experience is responsible for having replaced Heisenberg's operationally consistent matrix formulation of QM ---which made reference to the {\it intensive values} of observed spectral lines--- with Dirac's inconsistent axiomatic vectorial formulation ---in which the {\it ad hoc} introduction of single `clicks' and `measurements' within the theory has, in turn, also imposed the artificial distinction between `pure states' and `mixtures' \cite{deRondeMassri20a}. It was in fact Dirac who, through his reference to (binary) `certain measurements', introduced the essential confusion between a naive empiricist understanding of `actuality' as {\it hic et nunc} experience and a dogmatic metaphysical understanding of `actuality' as characterizing systems, states and properties. As we have discussed in a recent work \cite{deRonde20}, the so called ``measurement problem'' created in this process is nothing but the inconsistent conjunction of these two very harmful and oppressive perspectives of analysis. It was Dirac's imposition of single measurement outcomes which forced him to introduce the ``collapse'' of quantum superpositions, but it was his need to retain a (realist) physical discourse which also obliged him to ---anyhow--- make reference to ``particles''. 
\begin{quotation}
\noindent {\small``When we make the photon meet a tourmaline crystal, we are subjecting it to an observation. We are observing wether it is polarized parallel or perpendicular to the optic axis. The effect of making this observation is to force the photon entirely into the state of parallel or entirely into the state of perpendicular polarization. It has to make a sudden jump from being partly in each of these two states to being entirely in one or the other of them. Which of the two states it will jump cannot be predicted, but is governed only by probability laws.'' \cite[p. 9]{Dirac74}} 
\end{quotation}

Reinforcing classical metaphysical dogmatism, the rest of ``philosophical problems'' discussed in the literature have been grounded on the prejudice that the mathematical formalism of the theory of quanta should be explained in terms of the classical notions of locality, individuality, separability, distributivity, etc. In turn, this hypothesis has created a set of ``no-problems'' named after the failures of QM to describe classical systems. Thus, we have been left with the problems of quantum non-locality, quantum non-individuality, quantum non-separability, quantum non-distributivity, etc. The effect of these problems in the literature is twofold, not only they reinforce the reference to classical metaphysics ---which we already know is inadequate to describe the theory--- but ---more importantly--- they also prevent the development of a new (non-classical) conceptual scheme that would provide a deeper understanding about what the theory of quanta is really talking about. Exactly the same consequence is reached by the presupposition of a quantum to classical limit. Instead of concentrating the efforts in finding a consistent and coherent metaphysical scheme for understanding QM, the literature has focused in attempting to ``bridge the gap'' between the quantum mathematical formalism and our ``manifest image of the world''. The fact that no consistent representation has been reached has not stopped both physicists and philosophers from claiming that the phenomena is real ---at least, of course, ``For All Practical Purposes'' (FAPP).  

Analyzing things from a distance, it is quite paradoxical to observe that the same departure of QM from the classical representation of physics ---essentially described in terms of an {\it Actual State of Affairs} (see footnote 4) --- created the perfect environment for anti-realists not only to produce the most radical subversion of (realist) concepts but also to re-impose the same classical metaphysical scheme they had rightly criticized. It is, following {\it Corleone's Strategy}, that anti-realists were able to disguise their own debates with realist masks. The procedure followed by anti-realist has been to retain all the words applied in the realist system ---such as, `reality', `objectivity', `phenomena', `individual', `particle', etc.--- but detach them from their specific meaning, reference and content. Thus, while `reality' has become to signify `observation' or `fiction', `metaphysics' is understood today as `interpretation' or `narrative' (see \cite{deRonde20c}). In this process of disguising the meaning of words `objectivity' has been translated as `inter-subjectivity' and `measurement' ---depending on the need--- sometimes as the conscious act of observing a `click' and sometimes as the `physical interaction' of quantum and classical systems (see \cite{deRonde20}). Finally, the notion of `particle' ---detached from its metaphysical definition--- has been equated with a `click', an `event' or a `measurement outcome' (see \cite{deRonde20b}). It is through this radical re-signification of terms that anti-realists were able to create a monstrous labyrinth with no exit. Realists from all over the world have been welcomed to enter its gates at the reasonable price of repeating the following mantra: ``Empirical science is grounded in observations in the lab.'' Once inside the anti-realist facilities, realists are allowed to wonder freely creating and preaching their made-up stories to the few fellow companions willing to listen. Outside the gates, realists have been pictured ---depending on the needs--- either as ``naive believers'' in fictional stories created by themselves or as a ``dangerous fanatics'' who in their search for a single true story about reality could transform the perspectival nature of empirical science into a totalitarian regime. Most physicists, trained in an instrumentalist anti-philosophical fashion to ``shut up and calculate!'', do not even know about the existence of these facilities.\footnote{As Maximilian Schlosshauer \cite[p. 59]{Schlosshauer11} has recently described: ``It is no secret that a shut-up-and-calculate mentality pervades classrooms everywhere. How many physics students will ever hear their professor mention that there's such a queer thing as different interpretations of the very theory they're learning about? I have no representative data to answer this question, but I suspect the percentage of such students would hardly exceed the single-digit range.''} And why should they care? As they repeat to every open microphone: ``Physics has never been so successful!''\footnote{Lee Smoilin \cite{Smolin07} points to the existence of an unspoken major crisis which since the 1980s counts too many failures unrecognized to the rest of society: ``The current crisis in particle physics springs from the fact that the theories that have gone beyond the standard model in the last thirty years fall into two categories. Some were falsifiable, and they were falsified. The rest are untested ---either because they make no clean predictions or because the predictions they do make are not testable with current technology.''} Inside the walls, realists have been denied the possibility to choose a leader ---a single interpretation--- that would finally unite them and allow them to fight back. Anti-realists know very well their game: {\it Divide et impera.} Every realist is not only allowed but also encouraged by their captors to have her own preferences about which story to fight for. Fragmented and dispersed, realists wonder, sometimes alone, sometimes in small groups, through a map of madness fighting against each other in a hopeless search for their lost reality. In a field where words have lost their meaning, even anti-realists tour the facilities and sometimes have even fun creating their own interpretations ---of course--- detached from reality.

\section{Understanding Quantum Theory}

The first step to understand the theory of quanta is to abandon the anti-realist reference to ``empirical science'' which is grounded on a harmful combination of dogmatic metaphysics and naive empiricism. We have to stop taking for granted not only the observations of `clicks' as unproblematic {\it givens} of experience but also the idea that the metaphysics of particles is a necessary ingredient for QM. In this respect, the essential requirement to begin to discuss the physical meaning of QM is to revisit ---right from the start--- the meaning of physical theory itself and recognize the essential unity between physics and philosophy. As remarked by Schr\"odinger \cite[p. 12]{Schr54} already during the 1950s ---even before the rise of instrumentalism--- the separation between scientific and philosophical research ``produces the grotesque phenomenon of scientifically trained, highly competent minds with an unbelievably childlike --undeveloped or atrophied--- philosophical outlook.'' In this respect, the defense of philosophy from empirical scientists must be also regarded as the defense of physics from instrumentalism and anti-realism in general. We certainly agree with Tim Maudlin \cite{Maudlin19b} when he warns the instrumentalist about Standard Quantum Theory: ``No! You don't have a theory, you have a predictive recipe. And what we want, as a physicists what you should want is a theory [...]'' But in order to acknowledge this point it is also essential to admit that theories ---and in this point we disagree with Maudlin--- are not constructed from observations. Theories are developed from the attempt to produce representational formal-conceptual unified, consistent and coherent schemes of thought which are able to express experience adequately without any explicit reference to subjects. This is the true realist  quest which takes us ---once again--- back to the Greeks. As Heisenberg remarks:
\begin{quotation}
\noindent {\small ``What always distinguished Greek thought from that of all other peoples was its ability to change the questions it asked into questions of principle and thus to arrive at new points of view, bringing order into the colourful kaleidoscope of experience and making it accessible to human thought. [...] Whoever delves into the philosophy of the Greeks will encounter at every step this ability to pose questions of principle, and thus by reading the Greeks he can become practised in the use of the strongest mental tool produced by western thought.'' \cite[pp. 52-53]{Heis58b}} 
\end{quotation}

Sometimes the construction of theories, like in the case of classical mechanics, begins with physical concepts (e.g., space, time, continuum, particle, etc.) and requires the development of a mathematical formalism (e.g., infinitesimal calculus). But sometimes it is the other way around, like in the case of QM where we do possess a consistent mathematical formalism (i.e., Heisenberg's matrix formulation) but we still lack the concepts required in order to make sense of what is really going on. Whatever path is required, the definition of concepts is always essential to scientific understanding. Theories are not only composed of formal mathematical schemes, they are also constituted in an essential manner by conceptual schemes. And this is the main reason why we need to return to the original realist Greek tradition of scientific understanding. It is this tradition which has taught us that it is only theories which allow us to produce understanding of objective experience through unified, consistent and coherent representations. The question which guides the realist program has been always the same:  What is the relation between the One and the Many? What is {\it the same} within {\it difference}? In Modern times this question was rephrased in a more specific technical terms: How can we produce objective-invariant representations which explain experience as detached from subjects and reference frames? This, and not the creation of narratives, is the {\it praxis} which must unite realists under a common goal. For like in all matters of life, it is never so important what you actually belief or what you ---even--- choose to say in public, what is truly important is what you actually do.

\section*{Acknowledgements} 

I wish to thank Raimundo Fern\'andez Mouj\'an, Mat\'ias Graffigna, Axel Eljatib and Sergio Tonkonoff for discussions on subjects related to this paper. This work was partially supported by the following grants: the Project PIO CONICET-UNAJ (15520150100008CO) ``Quantum Superpositions in Quantum Information Processing'', UNAJ INVESTIGA 80020170100058UJ.


\begin{thebibliography}{9999}

\bibitem{Albert12} Albert, D., 2012, {\it Big Think Interview With David Albert}, https://www.youtube.com/watch?v=UNpLfXOfzZ8.

\bibitem{Albert19} Albert, D., 2019, ``David Albert on Quantum Measurement and the Problems with Many-Worlds'', {\it Sean Carroll's Mindscape Podcast (Episode 36)}, https://www.preposterousuniverse.com

\bibitem{BuenoVickers14} Bueno, O. $\&$ Vickers, P., 2014, ``Is science inconsistent?'', {\it Synthese}, {\bf 191}, 2887-2889.

\bibitem{Cabello17} Cabello, A., 2017, ``Interpretations of quantum theory: A map of madness'', in {\it What is Quantum Information?}, pp. 138-143,  O. Lombardi, S. Fortin, F. Holik and C. L\'opez (eds.), Cambridge University Press, Cambridge.

\bibitem{Carroll12} Carroll, S., 2012, Discussion surrounding the question, ``What is real?'' at the {\it Moving Naturalism Forward workshop},  https://www.preposterousuniverse.com/naturalism2012/

\bibitem{Carroll14} Carroll, S., 2014, ``Physicists Should Stop Saying Silly Things about Philosophy'', Personal Blog (Posted on June 23, 2014), https://www.preposterousuniverse.com/blog/2014/06/23/physicists-should-stop-saying-silly-things-about-philosophy/.

\bibitem{Carroll20} Carroll, S., 2020, ``A Brief History of Quantum Mechanics - with Sean Carroll'', {\it The Royal Institution}, https://www.youtube.com/watch?v=5hVmeOCJjOU

\bibitem{Cartwright83} Cartwright, N., 1983, {\it How the laws of physics lie}, Clarendon Press, Oxford.

\bibitem{Cordero14} Cordero, N.L., 2014, {\it Cuando la realidad palpitaba}, Biblos, Buenos Aires.

\bibitem{Crombie53} Crombie, A.C., 1953, {\it  Augustine to Galileo. The History of Science A.D. 400-1650.}, Harvard University Press, Massachusetts. 

\bibitem{PS} Curd, M. $\&$ Cover, J. A., 1998, {\it Philosophy of Science. The central issues}, Norton and Company (Eds.), Cambridge University Press, Cambridge.

\bibitem{daCostaFrench90} da Costa, N.C.A. $\&$ French, S., 1990, ``The Model-Theoretic Approach in the Philosophy of Science'', {\it Philosophy of Science}, {\bf 57}, 248-265.

\bibitem{deRonde20} de Ronde, C., 2020, ``The (Quantum) Measurement Problem in Classical Mechanics'',  in {\it Probing the Meaning of Quantum Mechanics: Entanglement, Correlations and Measurement}, D. Aerts, J. Arenhart, C. de Ronde, G. Sergioli (Eds.), World Scientific, Singapore, in press. (quant-ph:2001.00241).

\bibitem{deRonde20b} de Ronde, C., 2020, ``Measuring Quantum Superpositions (Or, ``It is only the theory which decides what can be observed.'')'', preprint. (quant-ph:2007.01146).

\bibitem{deRonde20c} de Ronde, C., 2020, ``Quantum Theory Needs No `Interpretation' But `Theoretical Formal-Conceptual Unity' (Or: Escaping Ad\'an Cabello's ``Map of Madness'' With the Help of David Deutsch's Explanations)'', preprint. (quant-ph:2008.00321).

\bibitem{deRondeFM19} de Ronde, C. $\&$  Fern\'andez Moujan, R., 2019, ``Epistemological vs Ontological Relationalism in Quantum Mechanics: Relativism or Realism?'' in {\it Probing the Meaning of Quantum Mechanics: Information, Contextuality, Relationalism and Entanglement}, pp. 277-317, D. Aerts, M.L. Dalla Chiara, C. de Ronde and D. Krause (Eds.), World Scientific, Singapore. 

\bibitem{deRondeMassri18a} de Ronde, C. $\&$ Massri, C., 2018, ``The Logos Categorical Approach to Quantum Mechanics: I. Kochen-Specker Contextuality and Global Intensive Valuations.'', {\it International Journal of Theoretical Physics}, DOI: 10.1007/s10773-018-3914-0. 

\bibitem{deRondeMassri20a} de Ronde, C. $\&$ Massri, C., 2019, ``Beyond Purity and Mixtures in Categorical Quantum Mechanics'', submitted. (quant-ph:2002.04423)

\bibitem{Deutsch04} Deutsch, D., 2004, {\it The Beginning of Infinity. Explanations that Transform the World}, Viking, Ontario. 

\bibitem{Dirac74} Dirac, P.A.M., 1974, {\it The Principles of Quantum Mechanics}, 4th Edition, Oxford University Press, London.

\bibitem{Einstein65} Einstein, A., 1949, ``Remarks concerning the essays brought together in this co-operative volume'', in {\it Albert Einstein. Philosopher-Scientist}, P.A. Schlipp (Eds.), pp. 665-689, MJF Books, New York. 

\bibitem{FM20} Fern\'andez Moujan, R., 2020, ``Greek philosophy for quantum physics. The return to the Greeks in the works of Heisenberg, Pauli and Schr\"odinger'' in {\it Probing the Meaning of Quantum Mechanics}, D. Aerts, J. Arenhart, C. de Ronde and G. Sergioli (Eds.), World Scientific, Singapore, in press. 

\bibitem{Feynman63} Feynman, R.P., Leighton, R.B. $\&$ Sands, M., 1963, {\it Lectures on Physics, Volume 1}, McGraw Hill, New York. 

\bibitem{Feynman67} Feynman, R.P., 1967, {\it The Character of Physical Law}, Massachusetts Institute of Technology Press, Massachusetts. 

\bibitem{Feynman85} Feynman, R.P., 1985, {\it Surely you're joking Mr. Feynman!}, Norton, New York. 

\bibitem{Feynman99} Feynman, R.P., 1999, {\it The Pleasure of Finding Things Out}, Perseus Publishing, New York. 

\bibitem{Fine86} Fine, A., 1986, {\it The Shaky Game}, University of Chicago Press, Chicago.

\bibitem{Freire15} Freire Jr., O. 2015, {\it The Quantum Dissidents. Rebuilding the Foundations of Quantum Mechanics (1950-1990)}, Springer, Berlin. 

\bibitem{Hawking88} Hawking, S., 1988, {\it A Brief History of Time}, Bantam Dell Publishing Group, New York. 

\bibitem{Hawking11} Hawking, S., 2011, ``Philosophy is Dead'', {\it Google Zeitgeist}, https://www.youtube.com/watch?v=pdLdA8E1Oa0

\bibitem{Heis58} Heisenberg, W., 1958, {\it Physics and Philosophy}, World perspectives, George Allen and Unwin Ltd., London.

\bibitem{Heis58b} Heisenberg, W., 1958, {\it The Physicist's Conception of Nature}, Harcourt, New York. 

\bibitem{Heis71} Heisenberg, W., 1971, {\it Physics and Beyond}, Harper \& Row, New York.

\bibitem{Heis73} Heisenberg, W., 1973, ``Development of Concepts in the History of Quantum Theory'', in {\it The Physicist's Conception of Nature}, pp. 264-275, J. Mehra (Ed.), Reidel, Dordrecht.


\bibitem{Kaku12} Kaku, M., 2012, ``The Universe in a Nutshell (Full Presentation)'', {\it Big Think}, min 5.40 https://www.youtube.com/watch?v=0NbBjNiw4tk

\bibitem{Krauss12} Krauss, M., 2012, ``Has Physics Made Philosophy and Religion Obsolete?'', {\it The Atlantic}, https://www.theatlantic.com/technology/archive/2012/04/has-physics-made-philosophy-and-religion-obsolete/256203/

\bibitem{Kuhn57} Kuhn, T., 1957, {\it  The Copernican Revolution: Planetary Astronomy in the Development of Western Thought}, Harvard University Press, Massachusetts.

\bibitem{Maudlin19} Maudlin, T., 2019, {\it Philosophy of Physics. Quantum Theory}, Princeton University Press, Princeton.

\bibitem{Maudlin19b} Maudlin, T., 2019, ``The Problem With Quantum Theory - Full Interview - Tim Maudlin'', {\it The Institute of Art and Ideas}, https://www.youtube.com/watch?v=hC3ckLqsL5M.

\bibitem{Meheus02} Meheus, J. (Ed.), 2002, {\it Inconsistency in science.} Kluwer Academic Publishers, Dordrecht.

\bibitem{Laplace02} Laplace, P.-S. 1902, {\it A Philosophical Essay on Probabilities}, John Wiley \& Sons, New York.


\bibitem{Pauli94} Pauli, W., 1994, {\it Writings on Physics and
Philosophy}, Enz, C. and von Meyenn, K. (Eds.), Springer,
Berlin.

\bibitem{Pigliucci09} Pigliucci, M., 2009, ``On the difference between science and philosophy'', https://rationallyspeaking.blogspot.com/2009/11/on-difference-between-science-and.html

\bibitem{Pigliucci12} Pigliucci, M., 2012, ``Lawrence Krauss: another physicist with an anti-philosophy complex'', https://rationallyspeaking.blogspot.com/2012/04/lawrence-krauss-another-physicist-with.html

\bibitem{Pigliucci14} Pigliucci, M., 2014, ``Neil deGrasse Tyson and the value of philosophy'', https://scientiasalon.wordpress.com/2014/05/12/neil-degrasse-tyson-and-the-value-of-philosophy/

\bibitem{Popper63} Popper, K.R., 1963,  {\it Conjectures and Refutations: The Growth of Scientific Knowledge}, Routledge Classics, London. 

\bibitem{Popper92} Popper, K., 1992, {\it The Logic of Scientific Discovery}, Routledge, New York. 

\bibitem{Rovelli07} Rovelli, C., 2007, {\it The First Scientist. Anaximander and his Legacy}, Westholme Publishing, Yardley.

\bibitem{Rovelli14} Rovelli, C., 2014, ``Why Talk to Philosophers? Part II.'', https://www.rotman.uwo.ca/why-talk-to-philosophers-part-ii/.

\bibitem{Schr54} Schr\"odinger, E., 1954, {\it Nature and the Greeks}, Cambridge University Press, Cambridge.

\bibitem{Schlosshauer11} Schlosshauer, M. (Ed.), 2011, {\it Elegance and Enigma. The Quantum Interviews}, Springer-Verlag, Berlin.

\bibitem{Smolin07} Smolin, L., 2007, {\it The trouble with physics. The rise of string theory, the fall of a science, and what comes next}, Mariner books, New York.

\bibitem{Smolin14} Smolin, L., 2014, ``Why Talk to Philosophers? Part III.'', https://www.rotman.uwo.ca/why-talk-to-philosophers-part-iii/.

\bibitem{VF80} Van Fraassen, B.C., 1980, {\it The Scientific Image}, Clarendon, Oxford.

\bibitem{VF91} Van Fraassen, B.C., 1991,  {\it Quantum Mechanics: An Empiricist View}, Clarendon, Oxford.

\bibitem{VF02} Van Fraassen, B. C., 2002, {\it The Empirical Stance}, Yale University
Press, New Haven.

\bibitem{Vernant06} Vernant, J.-P., 2006, {\it Myth and Thought among the Greeks}, Mariner books, New York.

\bibitem{Viereck} Viereck, G.S., 1930, {\it Glimpses of the Great}, Macauley, New York. 

\bibitem{Vickers13} Vickers, P., 2013, {\it Understanding inconsistent science}, Oxford University Press, Oxford. 

\bibitem{Vickers14} Vickers, P., 2014, ``Scientific Theory Eliminativism'', {\it Erkenntnis}, {\bf  79}, 111-126.

\bibitem{Weinberg88} Weinberg, S., 1988, Interview: ``Steven Weinberg on the Relationship Between Scientific Inquiry and Everyday Living'', September 23, 1988, {\it A World of Ideas}, https://billmoyers.com/series/a-world-of-ideas/

\bibitem{Weinberg93} Weinberg, S., 1993, {\it Dreams of a Final Theory: The Search for the Fundamental Laws of Nature}, Vintage, New York. 

\bibitem{Weinberg15} Weinberg, S., 2015, {\it To Explain the World: The Discovery of Modern Science}, Harper Collins, New York. 

\end{thebibliography}
\end{document}